\documentclass[graybox, natbib, authoryear, round]{svmult}

\usepackage{xcolor}
\usepackage{mathptmx}       
\usepackage{helvet}         
\usepackage{courier}        
\usepackage{type1cm}        
\usepackage{makeidx}         
\usepackage{graphicx}        
\usepackage{multicol}        
\usepackage[bottom]{footmisc}
\usepackage{hyperref}        
\usepackage{soul}            
\hypersetup{colorlinks=true, citecolor=black, urlcolor=blue}
\makeindex             

\def\etal{{et\,al.}}
\def\xray{\hbox{X-ray}}
\def\chandra{{\itshape Chandra\/}}
\def\ltsima{$\; \buildrel < \over \sim \;$}
\def\simlt{\lower.5ex\hbox{\ltsima}}
\def\gtsima{$\; \buildrel > \over \sim \;$}
\def\simgt{\lower.5ex\hbox{\gtsima}}
\def\msun{$M_\odot$}

\newcommand{\rmn}[1] {{\rm #1}}

\begin{document}
\title*{X-ray Binaries in External Galaxies}
\author{Marat Gilfanov \thanks{corresponding author}, Giuseppina Fabbiano, Bret Lehmer and Andreas Zezas
}
\institute{
Marat Gilfanov
\at
Max-Planck-Institute for Astrophysics, Garching, Germany and Space Research Institute, Moscow, Russia,
\\\email{gilfanov@mpa-garching.mpg.de}
\and
Giuseppina Fabbiano 
\at  Harvard-Smithsonian Center for Astrophysics (CfA), Cambridge MA, USA,
\\\email{gfabbiano@cfa.harvard.edu}
\and 
Bret Lehmer \at University of Arkansas, USA 
\\\email{lehmer@uark.edu}
\and
Andreas Zezas
\at University of Crete, Crete, Greece, 
\\\email{azezas@physics.uoc.gr}}
%
%
\maketitle
\abstract{X-ray appearance of normal galaxies is mainly determined by X-ray binaries powered by accretion onto a neutron star or a stellar mass black hole. Their populations  scale with the star-formation rate and stellar mass of the host galaxy and their X-ray luminosity distributions show a significant split between star-forming and passive galaxies, both facts being consequences of the dichotomy between high- and low-mass X-ray binaries. Metallicity, IMF and stellar age dependencies, and dynamical formation channels add complexity to this  picture. The numbers of high-mass X-ray binaries observed in star-forming galaxies indicate quite high probability for a massive star to become an accretion powered X-ray source once upon its lifetime. This explains the unexpectedly high contribution of X-ray binaries to the Cosmic X-ray Background, of the order of $\sim 10\%$, mostly via X-ray emission of faint star-forming galaxies located at moderate redshifts which may account for the unresolved part of the CXB. Cosmological evolution of the $L_X-\rmn{SFR}$ relation can make high-mass X-ray binaries a potentially significant factor  in  (pre)heating of intergalactic medium in the early Universe.
}


\section{Introduction}
\label{sec:intro}

X-ray binaries (XRBs) are binary stellar systems composed of a relativistic compact object -- a neutron star (NS) or a black hole (BH), and a stellar companion. They are powered by accretion of matter from the donor star onto the compact object \citep{ss73} and are the most common luminous compact X-ray sources in the Milky Way (see \citealt{lewin95, lewin06, gilfanov2010} for a review). While some luminous XRBs were detected in the early days of X-ray astronomy in Local Group galaxies, it was only after the deployment of the first imaging X-ray telescope, the Einstein Observatory in 1978, that their presence as ubiquitous populations of X-ray sources in galaxies could be established  (see \citealt{fabbiano89} for a review). A number of individual XRBs were resolved with Einstein in the more nearby galaxies \citep[e.g.][]{vsp79, lvsp83a}, leading to the first attempts at characterizing these populations with luminosity functions (see \citealt{fabbiano88} for a M81 – M31 comparison). These individual detections included a new class of very luminous X-ray sources  emitting in excess of the Eddington luminosity limit of a $\sim 10$ solar mass accreting object \citep{lvsp83}, now known as ultraluminous X-ray sources (ULXs). 

X-ray emission was detected from over 230 galaxies of all morphological types covered by the Einstein field of view, resulting in the publication of the Einstein Catalog of Galaxies \citep{fabbiano92}. Studies of the Local Group galaxies led to the realization that different XRB populations reside in galaxies  \citep[e.g.][]{lvsp83a, helfand1984}, akin to those in the Milky Way \cite[cf][]{grimm02}.
For more distant galaxies, Einstein  and, a decade later, ROSAT Observatory, could not yet resolve compact X-ray populations to the sufficient depth and detail, however, meaningful correlations were found between integrated X-ray emission and various multiwavelength tracers such as optical, near- and far-infrared and radio emission \citep{david1992, fabbiano02}.
These discoveries led to formulation of many elements of the overall picture of X-ray binary populations in galaxies. 
It was realised, in particular,  that young XRB population, identified with high-mass X-ray binaries (HMXBs), is prevalent in the arms of spiral galaxies and in late-type systems with young stellar population and more intense star formation rate. An older population of low-mass X-ray binaries (LMXBs) is instead found in the older galaxy disks and in the bulges. 

Following the demise of the Einstein Observatory, the study of galaxies was continued with other X-ray telescopes (ROSAT, ASCA, XMM-Newton). However, the real new breakthroughs in the study of the XRB populations have only occurred thanks to the deployment of the Chandra X-ray Observatory \citep{weisskopf00}. With its  sub-arcsecond angular resolution, Chandra  has revolutionized this field, leading to the widespread detection of rich populations of XRBs in galaxies well outside the Local Group (see reviews \citep{fabbiano, fabbiano19, gilfanov2004c}). 

In this Chapter we give an overview of the contemporary understanding of populations of XRBs in external galaxies and of their properties. The following discussion is mostly based on the results of Chandra observations. For the comprehensive review of earlier findings we refer the reader to  \citep{fabbiano89}.

\section{High- and low-mass X-ray binaries}
\label{sec:hmxb_lmxb}

Depending on the mass of the  optical companion, X-ray binaries are broadly divided into two classes -- high-  and low- mass X-ray binaries (HMXBs and LMXBs),
separated by a thinly populated region between $\sim 1$ M$_\odot$ and $\sim 5$ M$_\odot$, where bright persistent X-ray sources are scarce, for the reasons discussed in Section \ref{sec:pop_synthesis}.
The difference in the mass of the donor star determines the difference in the formation time scales  of these two classes of X-ray binaries, which are governed by a combination of the nuclear timescale of the donor stars and the time required for the onset of mass transfer from the donor star to the compact object. 

In the case of an HMXB, this  timescale is determined by the nuclear evolution  of the massive donor star. Correspondingly, they are X-ray bright within $\sim 100$ Myrs after formation of the binary system  \citep[e.g.][]{verbunt}. This is comparable to the characteristic timescale of the star-formation episode, therefore one may expect that the number of such systems in a galaxy is proportional to
its star-formation rate (SFR) \citep{stm78, grimm03, mineo12}:
\begin{equation}
 N_{\rm HMXB}, ~L_{\rm X,HMXB}\propto {\rm SFR}
\end{equation}
Evolution of primordial LMXBs is determined by the rate of loss of the orbital 
angular momentum of the binary system or by the nuclear evolution of the low-mass star, both of which, on the contrary to prompt HMXBs, are 
typically in the $\sim 1-10$ Gyrs range  \citep{verbunt}. Correspondingly, one may expect that their population 
integrates the long-term star-formation history of the host galaxy and scales with the total mass of its stars \citep{lmxb}:
\begin{equation}
N_{\rm LMXB}, ~L_{\rm X,LMXB} \propto {\rm M_*}
\end{equation}
LMXBs can be also formed dynamically in globular clusters and galactic nuclei which complicates this simple picture; this is  discussed in Section \ref{sec:dynamical}.

\section{X-ray scaling relations and luminosity functions}

\label{sec:xrb_scaling}

\subsection{Disentangling HMXB and LMXB populations in external galaxies}

High- and low-mass X-ray binaries scale, respectively,  with the SFR and stellar mass  of their parent stellar populations (Section \ref{sec:hmxb_lmxb}). They also share different evolutionary paths \citep[e.g.][]{Tauris_vdHeuvel}. For these reasons it is often useful to discriminate between the two classes in observations.  

With the exception of a few nearby galaxies \citep[for example Magellanic Clouds; e.g.][]{Antoniou2010, haberl2012, shtykovskiy2005a, shtykovskiy2005b} or a few detailed HST-based studies \citep[e.g.][]{Garofali2018},  it is extremely difficult to  determine the nature of the donor star in  galaxies located beyond $\sim 1$ Mpc, which would allow us to directly discriminate between high- and low-mass X-ray binaries. 

However, given that HMXBs are fueled by short-lived early-type OB stars, they become extinct after a few hundred million years after their formation. Therefore,  old stellar environments are populated by LMXBs. Similarly,  young stellar environments are dominated by HMXBs, (a) because of the much higher formation efficiency of HMXBs \citep[e.g.][]{Antoniou2019, mineo12} (Section \ref{sec:xrb_freqs}), and (b) because LMXBs did not have the time to form in environments much younger than $\sim$Gyr.  Therefore, we can use  estimates of relative fractions of old and young stellar populations in a galaxy to disentangle populations of HMXBs and LMXBs. A commonly used proxy for assessing the dominant XRB population is the specific SFR (sSFR) defined as the ratio of the SFR over the stellar mass at the same region of a galaxy. Generally, regions with sSFR higher than $\sim10^{-10}\rm{yr^{-1}}$ are dominated by HMXBs \citep{grimm03, mineo12, lehmer2010}, whereas  sSFR$\leq 10^{-11}\rm{yr^{-1}}$ would suggest an LMXB-dominated environment, although these thresholds can vary depending on the recent star-formation history of the galaxy.

\subsection{X-ray scaling relations.}
\label{sec:scaling_relations}

\begin{figure}[t]
\centering
\hbox{
\includegraphics[width=.5\textwidth]{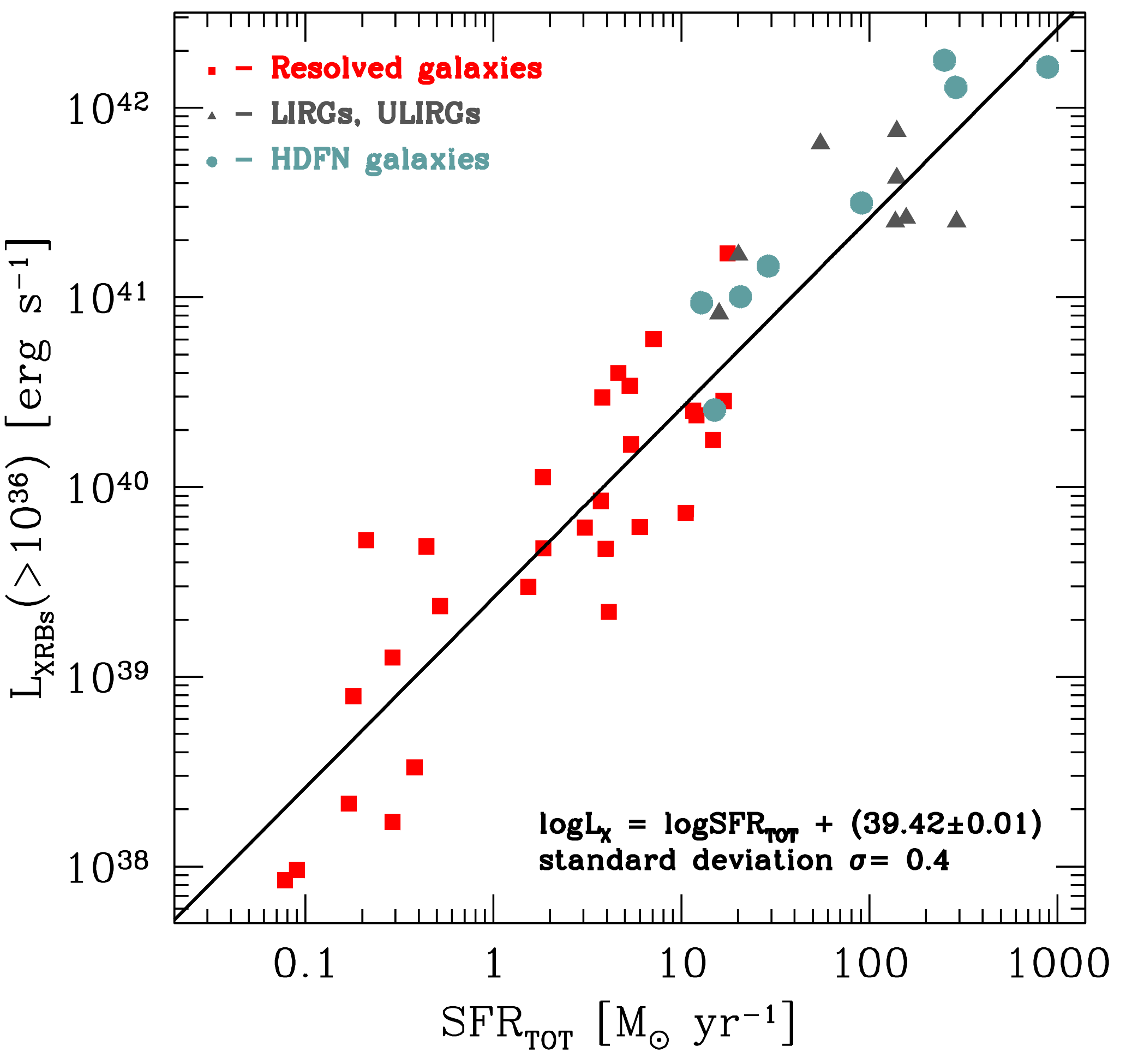}
\includegraphics[width=.5\textwidth]{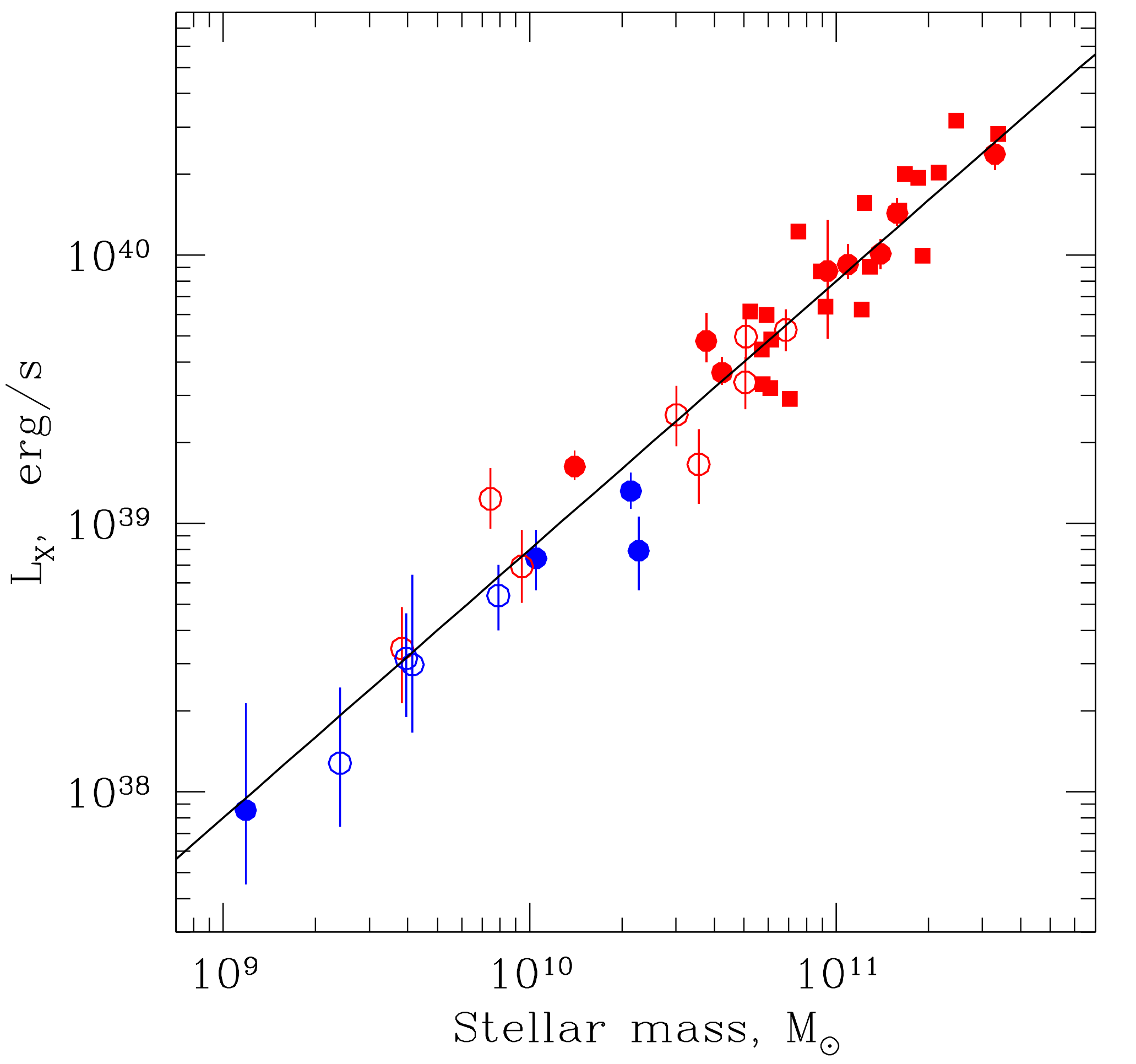}
}
\caption{Dependence of the total X-ray luminosity of X-ray binaries on the SFR (left panel) and stellar mass 
(right panel) of the host galaxy. Left panel shows star-forming galaxies, young stellar population of which is dominated by massive 
X-ray binaries; their population is roughly proportional to the SFR of the host galaxy. Right panel shows data for 
elliptical galaxies where star-formation mostly stopped at least several Gyrs ago and only low-mass X-ray binaries are left. 
Their population is determined by the total stellar mass of the host galaxy. Solid lines show approximation of the data by the 
linear laws. In the left panel we also show the data for ultra-luminous infrared galaxies (ULIRGs, triangles) and star-forming galaxies from the Chandra Deep Fields. 
These galaxies are not resolved by Chandra, therefore the total luminosity is shown, including contribution of faint unresolved compact 
sources and diffuse emission. Adapted from   \citep{lmxb,mineo12,zhang12}. 
}
\label{fig:lx}
\end{figure}

\begin{figure}
\centering
\hbox{
\includegraphics[width=.6\textwidth]{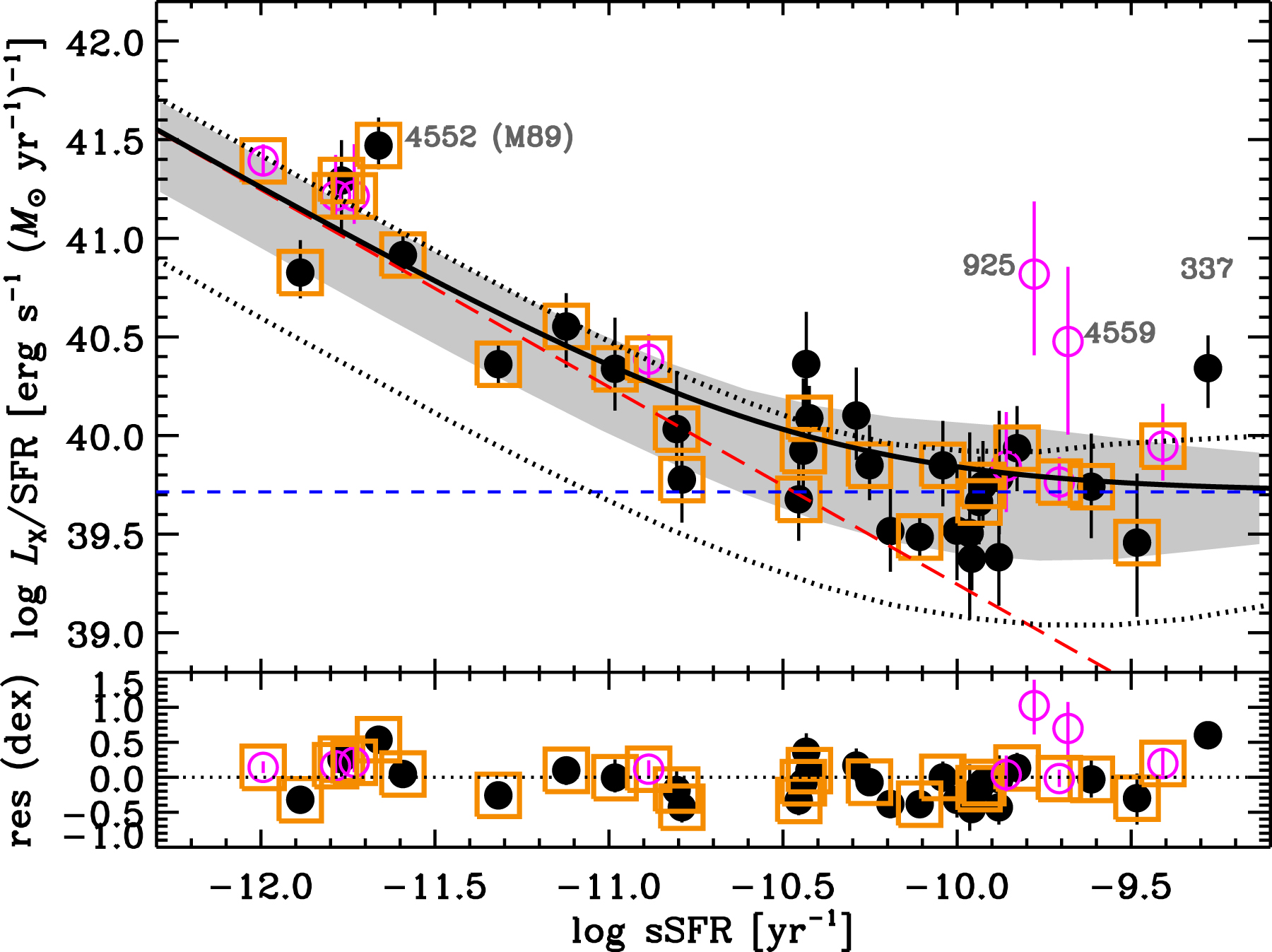}
\hspace{0.3cm}
\begin{minipage}[b]{0.37\textwidth}
\includegraphics[width=\textwidth]{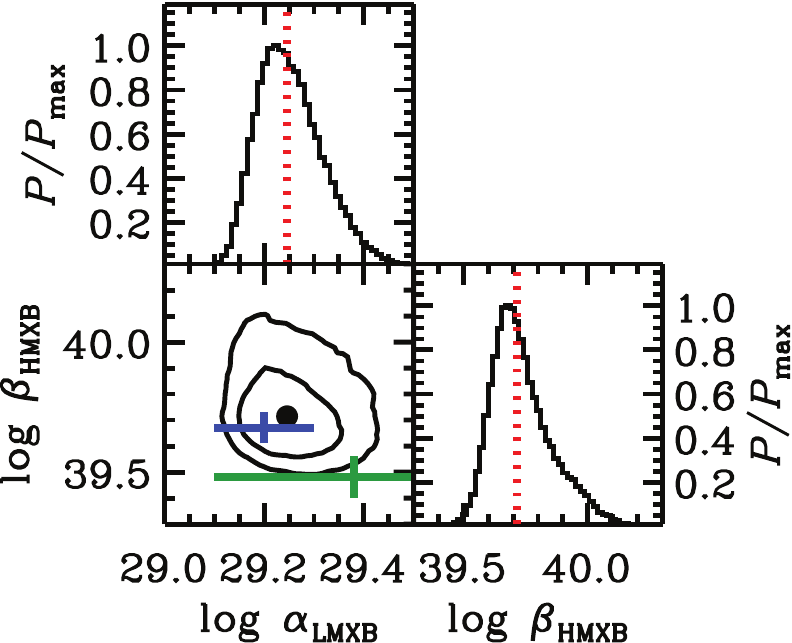}
\vspace{1.2cm}
\end{minipage}
}
\caption{{\em Left:} Dependence of the total X-ray luminosity per unit SFR on the sSFR of the host galaxy. The red-dashed line indicates the contribution from LMXBs, while the blue-dashed line indicates the contribution from HMXBs.   Galaxies with sSFR greater than ($\sim 10^{-10.5}\,\rm{M_{\odot}\,yr^{-1}\,M_{\odot}^{-1}}$) are dominated by HMXBs while at lower values of sSFR the contribution of LMBXs becomes increasingly more important. The gray shaded region and the dotted lines delineate the 1$\sigma$ scatter resulting from uncertainties in the  XLF of the resolved X-ray binaries for galaxies with median mass of $2\times10^{10}$\msun, and  $3\times10^{9}$\msun.
{\em Right:} Comparison of scaling parameters $\alpha_{\rm LMXB}$ and $\beta_{\rm HMXB}$  from eq. (\ref{eq:xrv_scaling_lehmer}) derived in \cite{lehmer2019} (MCMC contours) with those from eqs. (\ref{eq:hmxb}) and (\ref{eq:lmxb}) \citep{lmxb, mineo12, zhang12} (blue cross with corresponding $1\sigma$ errors) after differences in assumed IMF and SFR, $M_*$ proxies are taken into account. The green cross shows the Chandra Deep Field-South independent estimates from \cite{lehmer16}.
See \cite{lehmer2019} for more details. Adapted from \cite{lehmer2019}.
}
\label{fig:lx_sSFR}
\end{figure}

Chandra X-ray observatory presented an opportunity to observe compact sources in  galaxies located at distances up to $\sim 30-50$ Mpc (and more for the brightest sources)
in a nearly confusion-free regime, to  measure their luminosity functions and total luminosities of  different (sub)populations. Observations of a large number ($\sim$ hundred) of nearby galaxies have demonstrated that populations of LMXBs and 
HMXBs in a galaxy scale  proportionally to its stellar mass and SFR respectively (Fig.\ref{fig:lx}): 
\begin{eqnarray}
\label{eq:hmxb}
L_{\rm X, HMXB}\approx 2.6\cdot 10^{39}\times {\rm SFR}~~~~~~~~~~
N_{\rm HMXB}\approx 10\times {\rm SFR}~~~~~ \\
\label{eq:lmxb}
L_{\rm X, LMXB}\approx 10^{39}\times \frac{\rm M_*}{10^{10} M_\odot}~~~~~~
N_{\rm LMXB}\approx 14\times \frac{\rm M_*}{10^{10} M_\odot}
\end{eqnarray}
where $L_{\rm X}$ is the total X-ray luminosity of X-ray binaries of the given type in the  0.5--8 keV energy band,  
$N_{\rm X}$ the number of X-ray binaries with luminosity exceeding  $L_{\rm X}\ge 10^{37}$ erg/s, SFR is the star-formation 
rate in  M$_\odot$/yr, and  M$_*$ is the stellar mass of the galaxy in solar units \citep{grimm03,mineo12,lmxb}. 
Broadly consistent results were obtained in several other independent studies \citep[e.g.][]{ranalli,colbert04,kim2004,lehmer2010, sazonov2017}. However, there is a caveat to keep in mind when comparing different scaling relations. The coefficients in eq.(\ref{eq:hmxb}), (\ref{eq:lmxb}) depend first of all on the methods used for estimating the stellar mass and SFR, their proxies and assumed shape of the initial mass function (IMF). This is illustrated by Fig. \ref{fig:lx_sSFR} and explained later in this section. Secondly, they depend on the particular samples of galaxies, their age and star-formation history composition, metallicity, globular cluster content, etc.  Various aspects of these dependencies are discussed in the following sections of this chapter.

The $L_X-{\rm SFR}$ relation is subject to a curious statistical effect, making it to appear steeper than linear and to have large dispersion at low SFR \citep{grimm03, gilfanov2004b}. This behavior is caused by the poor sampling of the bright luminosity end of the X-ray luminosity function (XLF) at low SFR and its  magnitude depends on the XLF shape, in particular its slope and position of the high luminoisty cut-off \citep{gilfanov2004b}. It is much more pronounced for HMXBs because of their relatively flat XLF and is insignificant for LMXBs which XLF is steep at the high luminosity end (Section \ref{sec:xlf}). Note that this effect is not seen in Fig.\ref{fig:lx} because of the observer bias \citep{mineo12}.

In order to account for the HMXB and the LMXB components simultaneously  Lehmer and collaborators \citep{lehmer2010} introduced a scaling relation of the form
\begin{equation}
\rm{L_{X} = \beta SFR + \alpha M_{*}}
\end{equation} 
An equivalent form in terms of specific SFR is also often used:
\begin{equation}
\rm{L_{X}/SFR = \beta + \alpha\times sSFR^{-1}}
\label{eq:xrv_scaling_lehmer}
\end{equation} 
\citet{lehmer2019} derived values of the scaling parameters: $\log \alpha$ [erg~s$^{-1}$~$M_\odot^{-1}$] = 29.25$^{+0.07}_{-0.06}$ and $\log \beta$ [erg~s$^{-1}$~($M_\odot$~yr$^{-1}$)$^{-1}$]  = 39.71$^{+0.14}_{-0.09}$. Their best fit to the data is shown in Fig. \ref{fig:lx_sSFR}. 
This  parameterization allows one to combine the contribution of the LMXB and HMXB components in a single formulation, and it is particularly useful for galaxies with low sSFR such as early-type spiral galaxies that have a significant LMXB component.   

The parameters $\alpha$ and $\beta$ in eq. (\ref{eq:xrv_scaling_lehmer}) characterising X-ray scaling relations for LMXBs and HMXBs are in an apparent disagreement with those in eqs. (\ref{eq:hmxb}) and (\ref{eq:lmxb}). The disagreement is not real, as discussed earlier in this section, it is caused by the differences in the assumed IMF and  proxies used for the SFR and stellar mass estimations in the different analyses.  When they are taken into account  \citep{lehmer2019}, the scaling relations from \cite{lmxb, mineo12, zhang12, lehmer2019} are fully consistent with with each other as it is illustrated in the right panel in Fig.\ref{fig:lx_sSFR}.

\subsubsection{Time dependence of HMXB population}
\label{sec:eta_hmxb}

Scaling relations in Fig. \ref{fig:lx} and \ref{fig:lx_sSFR} are drawn for quantities  integrated  over entire galaxies. As such, they represent a result of averaging  over a (typically large) number of star-forming regions. The details regarding ages, metallicities  and star-formation histories of the individual star-forming regions are smeared out and such scaling relations characterise  populations X-ray binaries globally, on the galactic scales. Spatially resolved analysis of  individual galaxies, on the other hand, can reveal a more detailed picture of evolution of the population of HMXBs with time  after their formation.

\begin{figure}
\hbox{\centering
\includegraphics[width=.5\textwidth]{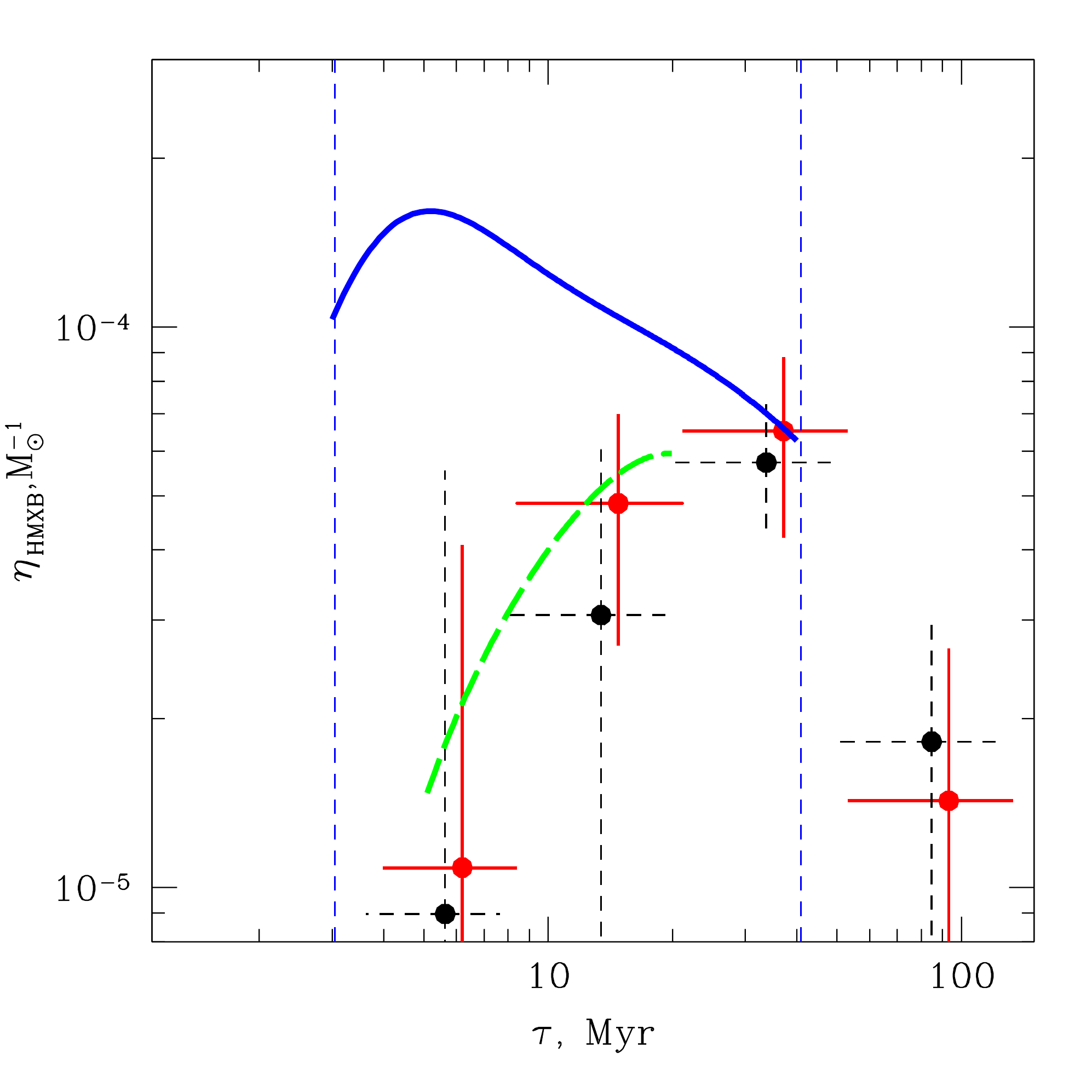}
\includegraphics[width=.5\textwidth]{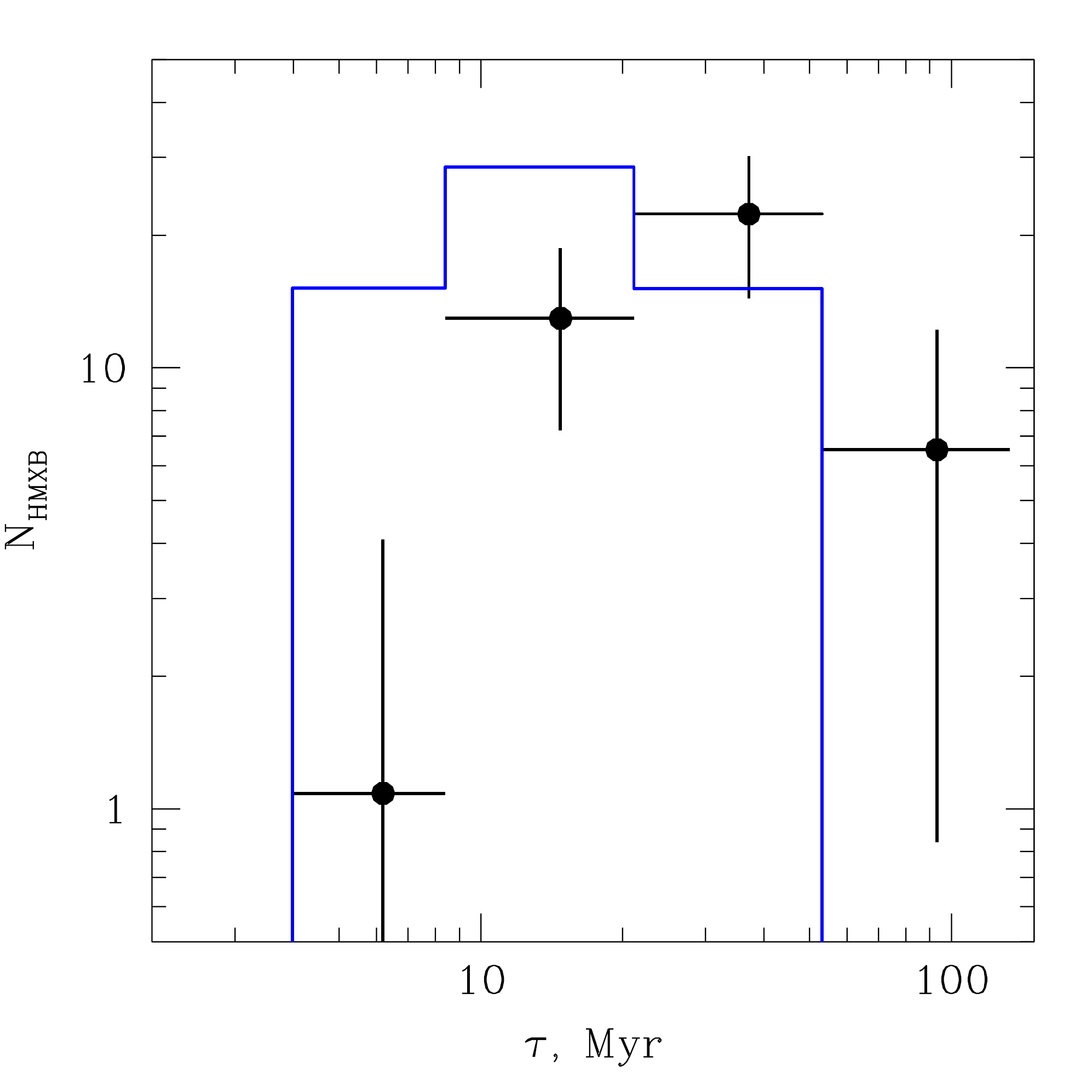}}
\vbox{\centering
\includegraphics[width=.6\textwidth]{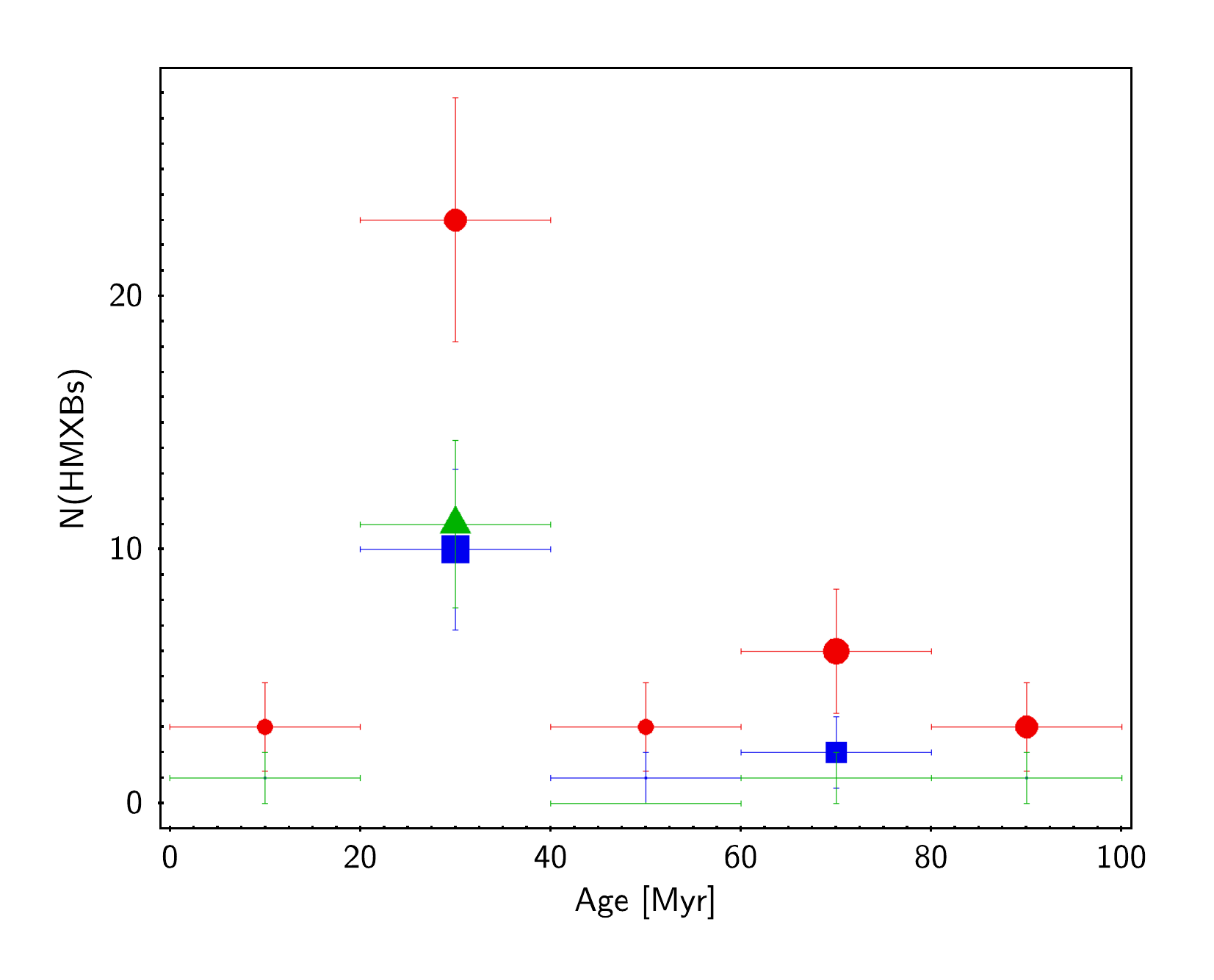}
}
\caption{
{\em Top-left panel:} Dependence of the HMXB number on the time elapsed since the 
star formation event. The solid and dashed
crosses were obtained using  different reconstructions of spatially resolved star-formation histories (see \cite{shtykovskiy2007b} for details). The solid
curve shows the type II supernova rate. The two vertical dashed
lines mark  formation times of the first
black hole and the last neutron star calculated in the standard theory of evolution of a single star. The dashed curve
represents the theoretical dependence of the number of Be/X systems with neutron stars 
from \cite{popov1998}. 
{\em Top-right panel:} The
age distribution of HMXBs in the SMC.  The solid histogram shows the distribution expected  if HMXB numbers followed the core collapse supernova rate. The top panels adapted from \cite{shtykovskiy2007b} 
{\em Bottom panel:} The
age distribution of HMXBs in the LMC. Adapted from \cite{Antoniou2016}
}
\label{fig:eta_hmxb}
\end{figure}

A particularly promising way is to compare the spatial distribution of HXMBs in a galaxy with its spatially resolved star-formation history or the  stellar population age. At present, this can be done only for the handful of the most nearby galaxies, and the results are  limited by the statistical quality of the available data, in particular, by the number of HMXBs and fairly poor sampling of their parameter (type, luminosity etc) space.

In the case of the Magellanic Clouds comparison of the X-ray binary populations with the star-formation history (SFH) of their parent stellar populations permitted the authors to reconstruct the HMXB formation efficiency $\eta_{\rm HMXB}(\tau)$ \citep[see definition in ][]{shtykovskiy2007b} and the  age distribution of HMXBs as shown in Fig. \ref{fig:eta_hmxb}. It was shown that there is a peak at their formation efficiency at $\sim30-60$\,Myr, consistent with increased formation of Be stars at the same timescales \citep{shtykovskiy2007b,Antoniou2010,Antoniou2016,Antoniou2019}. As it should have been expected, the peak in the population of HMXBs is  delayed with respect to the peak in the formation rate of compact objects (left and middle panels in Fig. \ref{fig:eta_hmxb}). The magnitude of the delay  is determined by the time required to form the compact object (mostly neutron stars in the case of HMXBs in Magellanic Clouds) and for the nuclear evolution of the secondary star.  These results give a formation rate of $\sim1$ X-ray binary per 200 observed stars of OB spectral types at ages of $\sim30-60$\,Myr, or equivalently $\sim9$ XRBs per $10^{6}\,\rm{M_{\odot}}$ of stars formed at a star-formation episode, which  agrees very well with the theoretical models, as one can see in the left panel in Fig. \ref{fig:eta_hmxb} (see also Fig. \ref{fig:lx_age}) 

Magellanic Clouds have fairly low star-formation rates and their X-ray populations are dominated by low luminosity sources \cite[e.g.][]{gilfanov2004b}. The $\eta_{\rm HMXB}(\tau)$ shown in Fig.\ref{fig:eta_hmxb}  describes X/Be systems which these galaxies are populated with, hence it peaks at $\sim$several tens of Myrs. Similar analysis for high-SFR galaxies (e.g. Antennae or Cartwheel) would reveal a complex picture of the formation efficiencies and time scales depending on X-ray luminosity. In particular one might expect to see that $\eta_{\rm ULX}(\tau)$ for ultra-luminous X-ray sources  peaks at much earlier times. However, such analysis would require to resolve stellar populations and to derive spatially resolved star-formation histories at distances of $\sim 10-100$ Mpc.

\subsubsection{Metallicity and age effects}

\begin{figure}
\centering
\includegraphics[width=.6\textwidth]{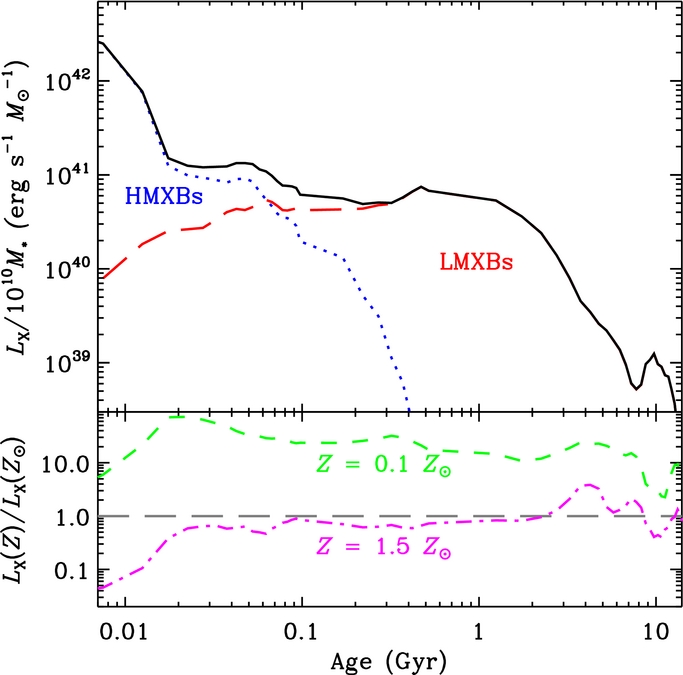}
\caption{The evolution of the X-ray luminosity of an initial stellar population of $10^{6}\,\rm{M_\odot}$ as a function of age. The bottom panel shows the ratio of the same model for sub-solar ($\rm{0.1\rm{Z_\odot}}$) and supersolar ($\rm{1.5\rm{Z_\odot}}$) metallicity with respect to the solar metallicity model. Adapted from \citep{fragos13}. }
\label{fig:lx_age}
\end{figure}

X-ray binary population synthesis models \citep[e.g.][]{fragos13, Linden2010} showed that the number of X-ray binaries, the shape of their luminosity function,  and their integrated X-ray luminosity evolve strongly as a function of age and metallicity (Fig. \ref{fig:lx_age}). The general trend is that as a stellar population ages the integrated X-ray luminosity of its X-ray binaries declines. In the case of metallicity, lower metallicity stellar populations are associated with  higher integrated X-ray luminosities.  

These predictions have been confirmed from studies of the X-ray luminosity or the number of X-ray binaries as a function of the age of the stellar populations \citep[e.g.][]{lehmer2019}. Similarly, studies of the effect of metallicity showed strong anti-correlation of the integrated X-ray luminosity of galaxies and their metallicity (see  discussion in section \ref{sec:evol} and Fig. \ref{fig:hz_scaling}) \citep[e.g.][]{brorby16,prestwich2013,mapelli2009,lehmer16}. Such behaviour is attributed to the weaker stellar winds of low-metallicity stars, resulting in tighter orbits at the start of the X-ray emitting phase, and hence higher probability of systems undergoing Roche-lobe overflow   \citep[e.g.][]{Linden2010}. Metallicity dependence  is further discussed in Section \ref{sec:evolution} in the context of cosmic evolution of X-ray binary populations in galaxies.

\subsubsection{Sub-galactic scales}

Sub-galactic scales open a window into the complex picture  of evolving X-ray  populations (Section \ref{sec:eta_hmxb}) however require adequate angular resolution and sensitivity to properly resolve optical and X-ray populations. At present, beyond the Local Group, one can only study behaviour on $\sim$few kpc scales of integrated (unresolved) X-ray luminosity and various SFR and mass proxies. 

Recent studies of the X-ray luminosity, stellar mass and SFR scaling relations in sub-galactic scales as small as 1\,$\rm{kpc^{2}}$ showed that unresolved X-ray luminosity  follows the same qualitative trends as the galaxy-wide scaling relations \cite{kouroump2020}. For larger sub-galactic regions of the $\sim 3-4$ kpc size and SFR in excess of $\sim 10^{-2}$ M$_\odot$yr$^{-1}$, correlations of $L_X-{\rm SFR}$ converge to the integrated galactic emission relations. In regions of smaller size and/or with extremely low SFR ($<10^{-2}$\,$\rm{M_{\odot}\,yr^{-1}}$)  the X-ray luminosity SFR relation shows a flattening which can be attributed to the contribution of the increasingly important LMXB component and, possibly, intrinsically fainter X-ray populations (e.g. active binaries and cataclismic variables). In addition these sub-galactic relations show significant scatter which is the result of stochastic effects (see Section \ref{sec:scaling_relations}) and variations of the stellar populations between different regions of a galaxy.

\subsection{X-ray luminosity functions}
\label{sec:xlf}

\begin{figure} 
\centering
\hbox{
\includegraphics[width=.5\textwidth]{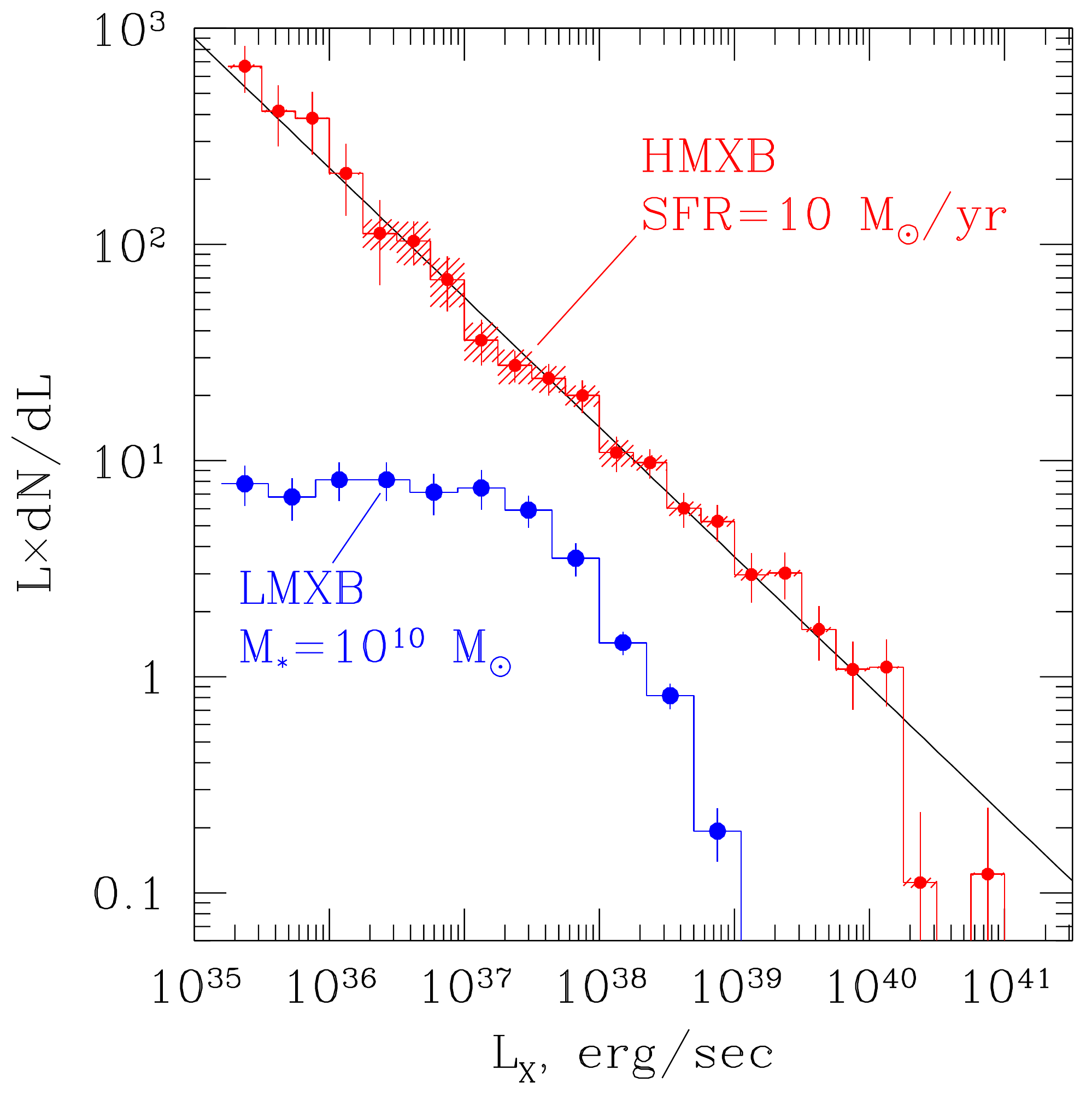}
\includegraphics[width=.5\textwidth]{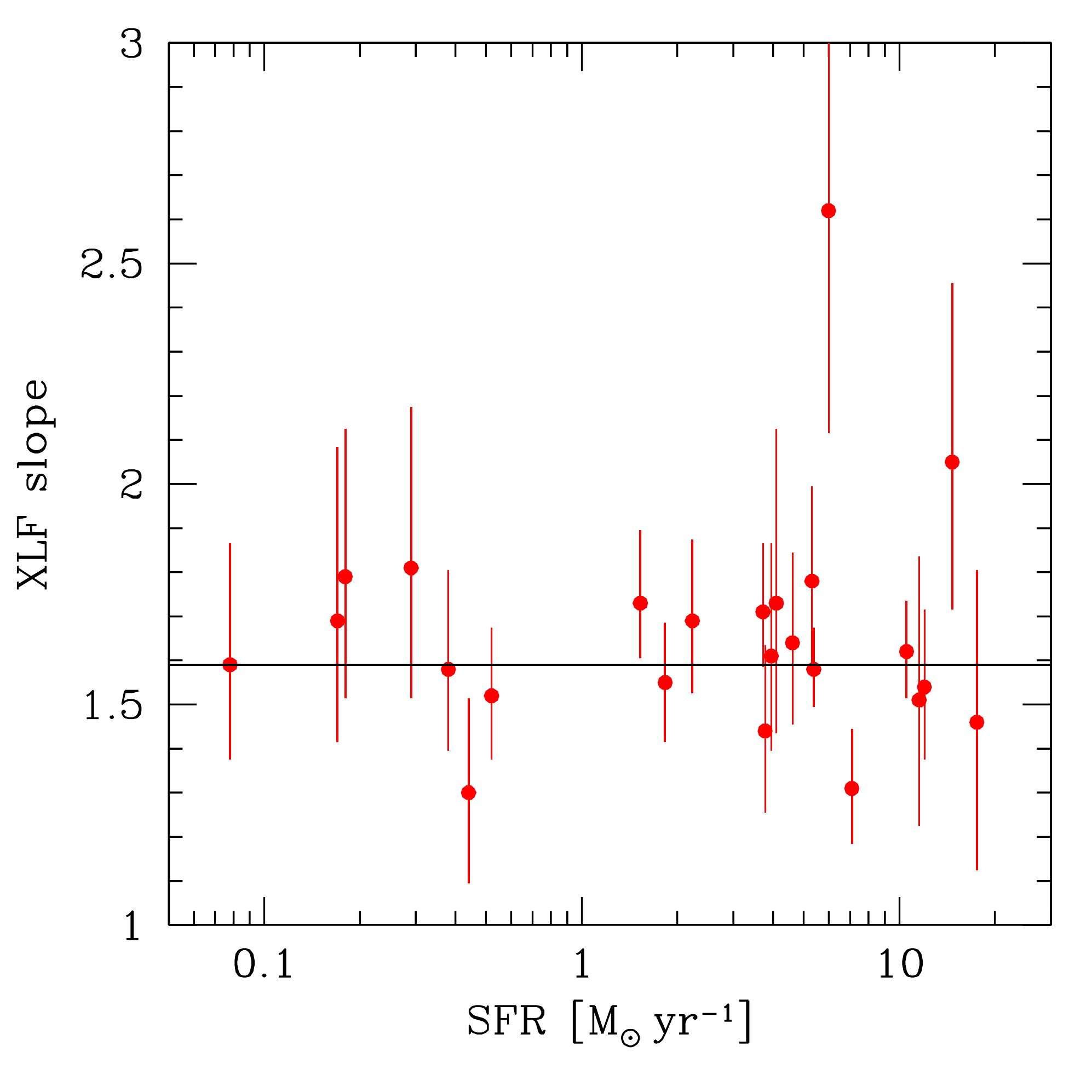}
}
\caption{{\em Left:} Average X-ray luminosity functions of of compact X-ray sources in star-forming (marked "HMXB") and elliptical (marked "LMXB") galaxies. In star-forming galaxies high-mass X-ray binaries dominate, whereas in old elliptical galaxies the dominant component of X-ray populations are low-mass X-ray binaries. Luminosity functions are normalized to star-formation rate of SFR$=10$ M$_\odot$/yr and stellar mass of $M_*=10^{10}$ M$_\odot$ respectively. Based on results of   \cite{lmxb} and \cite{mineo12}. {\em Right:}  XLF slopes for individual star-forming galaxies from the sample of Mineo et al. \citep{mineo12} plotted against the SFR. The solid horizontal line shows the slope of the mean HMXB XLF shown in the left panel. Adapted from \cite{mineo12}.}
\label{fig:xlf}
\end{figure}

Chandra observations of many nearby galaxies showed that  X-ray luminosity functions of their compact 
X-ray sources have quite similar shape (Fig.\ref{fig:xlf}, right panel), with much stronger variations in normalization than in shape \citep{grimm03, lmxb, mineo12}. It was also immediately noticed that shapes of XLFs  are different for X-ray populations in star-forming and elliptical galaxies, i.e. for high- and low- mass X-ray binaries (Fig.\ref{fig:xlf}, left panel). In agreement with the  scaling relations  (Section \ref{sec:scaling_relations}), normalization of the HMXB and LMXB luminosity functions scale proportionally to the star-formation rate and stellar mass of the host galaxy correspondingly. These  Chandra findings led to the conclusion that  luminosity distributions of X-ray binaries can be described, to the first approximation,  by the universal luminosity functions of HMXBs and LMXBs \citep{grimm03, lmxb} (Fig.\ref{fig:xlf}). As time went on and  Chandra  observations sampled  the parameter space of galaxies and also their depth increased, this picture was refined,  with  age, metallicity, globular cluster content  playing their roles in shaping the luminosity distributions of HMXBs and LMXBs.

\begin{figure} 
\centering
\hbox{
\includegraphics[width=.5\textwidth]{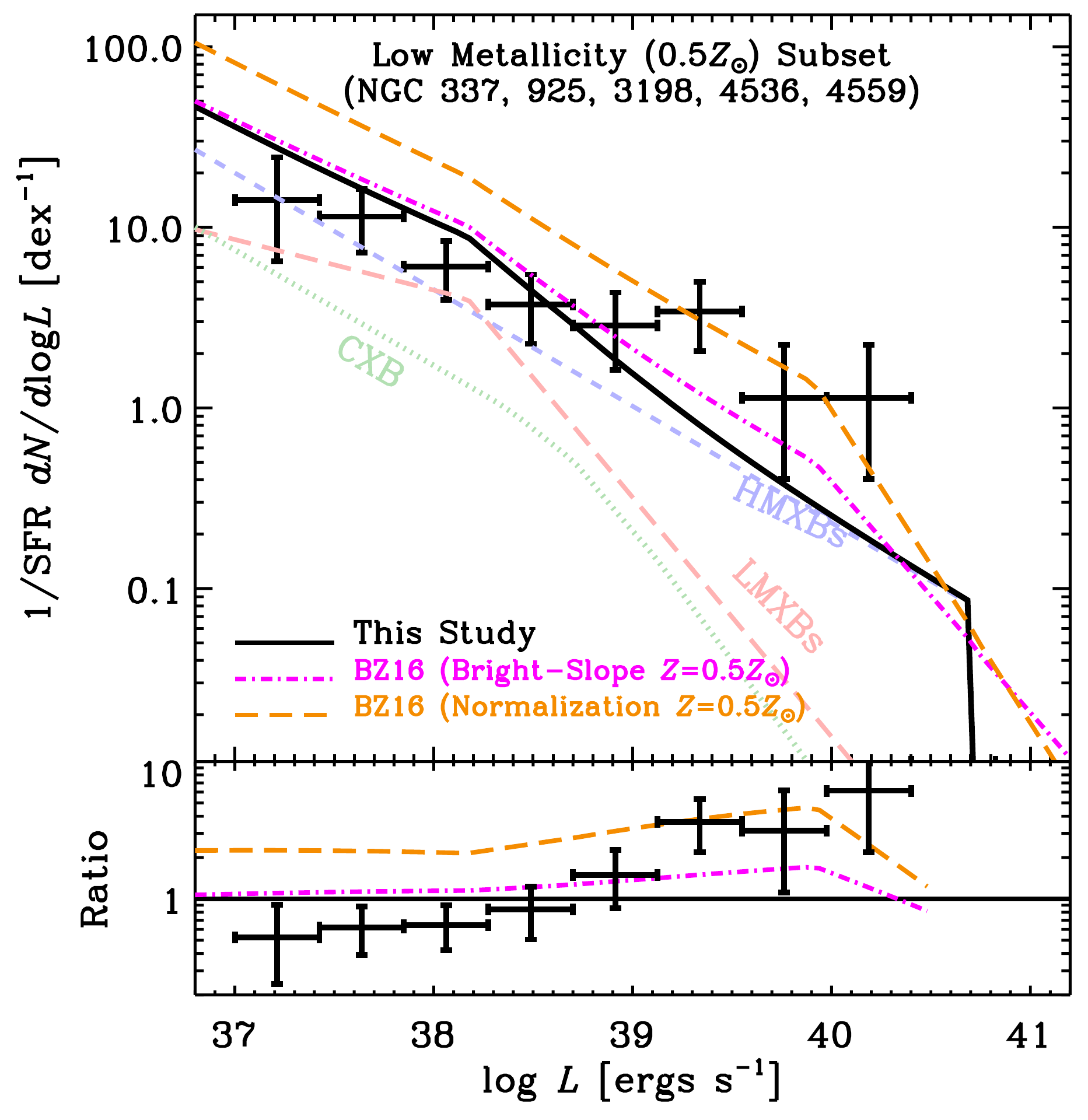}
\includegraphics[width=.47\textwidth]{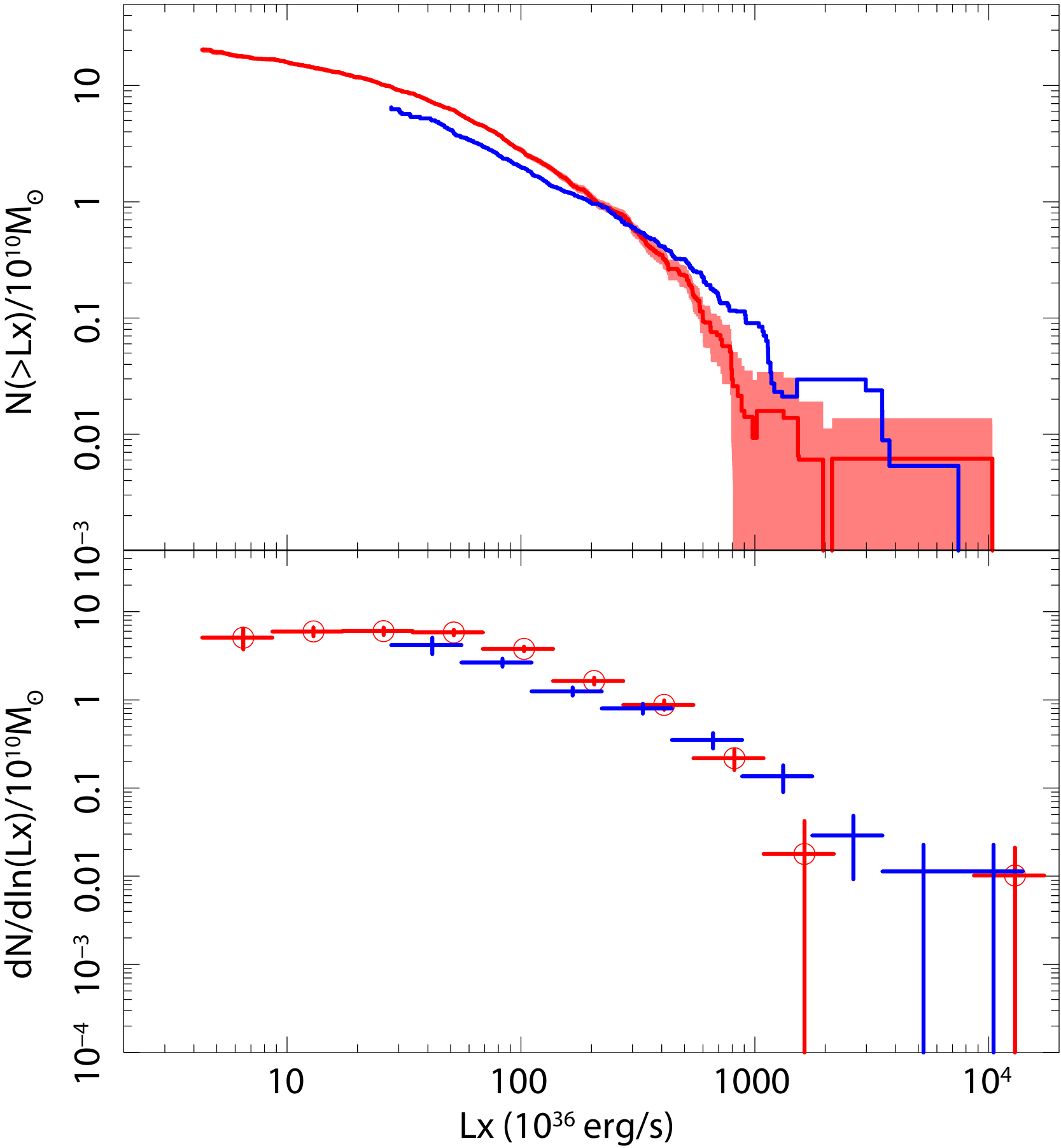}
}
\caption{{\em Left:} SFR-normalized incompletness-corrected  total XLF for  five low metallicity star-forming galaxies (NGC~337, 925, 3198, 4536, and 4559), which have metallicities of $\approx$0.5~$Z_\odot$.  The black curve
shows the global model \cite{lehmer2019} prediction for this population, including HMXB, LMXB, and CXB contributions. Enhancements in the $L \simgt 10^{39}$ erg/s source population are clearly observed. 
The bottom panel shows the ratio  of data and  population synthesis models with respect to the best-fit global model prediction. Adapted from \cite{lehmer2019}
{\em Right:} XLFs of LMXBs in young and old galaxies
in cumulative (upper panel) and differential (lower panel) forms. See \cite{zhang12} for description of the sample and further details. The data for old galaxies 
(red) is marked by circles in the lower panel and is 
surrounded by the shaded area showing  the $1\sigma$ Poissonian uncertainty in the upper panel. 
Statistical uncertainty for young galaxies has comparable amplitude. Adapted from \cite{zhang12}.
}
\label{fig:xlf2}
\end{figure}

These refinements added  complexity and detail to  our understanding of the XLFs of X-ray binaries, but have not changed the fact that the shapes of the  XLFs of HMXBs and LMXBs are qualitatively different (Fig.\ref{fig:xlf}). The 
difference is mainly caused by the difference in the mass transfer regime in high and low- mass X-ray binaries (see more detailed discussion in Section \ref{sec:pop_synthesis}). Indeed, in the majority of  the former the compact object accretes material from the wind of the massive donor star. Their the luminosity function is determined primarily by the mass distribution of the donors in high-mass X-ray binaries 
\citep[e.g.][]{postnov03} which leads to the formation of the observed power law luminosity distribution \citep{grimm03,mineo12}:
 \begin{equation}
\frac{dN_{\rm HMXB}}{dL_{\rm X}}\propto {\rm SFR}\times  L^{-1.6}
\label{eq:xlf}
\end{equation}
In the case of low-mass X-ray binaries, on the contrary, the mass transfer occurs via donor star Roche lobe overflow 
through the inner Lagrangian point of the binary system and the X-ray  luminosity function of these systems is determined by the orbital parameter distribution of semi-detached binary systems in the galaxy. This leads to formation of the luminosity distribution of a complex shape, with two breaks at $\log L_X\sim 38.5$ and $\log L_X\sim 37-37.5$ \citep{lmxb, kim10}. 
The first break is located near Eddington luminosity of the neutron star and is likely related with the existence of the luminosity limit for an accreting neutron star -- compact objects in more luminous systems are black holes whose occurrence rates in the population are smaller. The nature  of the second break is still unclear.

Using large sample of nearby normal galaxies from the Chandra archive, Lehmer and collaborators \citep{lehmer2019} analysed XRB populations on sub-galactic scales and built a global  XLF model accounting for both types of X-ray binaries and parameterised via SFR and stellar mass of the host galaxy. They found clear transition in XLF shape and normalization per SFR from the almost “pure” HMXB XLF at  sSFR$\simgt 5\cdot 10^{-10}$ yr$^{-1}$ to the nearly pure LMXB XLF at sSFR$\simlt 10^{-12}$ yr$^{-1}$. The large number of sources (over $\sim 4400$) in their sample permitted them to accurately describe  more subtle XLF features. 
Also, they found statistically significant evidence that the HMXB XLF in low-metallicity ($\approx 0.5 Z_\odot$) galaxies contains an excess of high luminosity $L_X\simgt 10^{39}$ erg/s sources compared to the global average HMXB XLF, which has a median metallicity $\approx Z_\odot$ (left panel in Fig.\ref{fig:xlf2}). This result is in line with  findings that the integrated X-ray luminosity per SFR is anticorrelated with metallicity \citep[e.g.][]{basu13, brorby16} and with prediction of the population synthesis modeling (Section \ref{sec:pop_synthesis}).

Thanks to their long evolution time scale, in the Gyrs range, time dependence of LMXB populations can be probed with the currently available data.  Some dependence on the stellar age is natural to expect, it is predicted in  population synthesis modeling (Section \ref{sec:pop_synthesis}) and detected in observations \citep{zhang12, lehmer2019}. There is  clear evolution of the LMXB XLF with age -- younger galaxies have more bright sources and fewer faint ones per unit stellar mass (right panle in Fig. \ref{fig:xlf2}). The XLF of LMXBs in younger galaxies appears to extend significantly beyond $\sim 10^{39}$ erg/s. Such bright sources seem to be less frequent in older galaxies. 

A natural question is to what degree the variability of the X-ray binary populations affects the shape of the X-ray luminosity function of a given galaxy. Systematic monitoring campaigns on a couple galaxies (the Antennae, M81) showed that while individual sources show significant variability the shape of their X-ray luminosity functions is remarkably stable \cite{zezas07}. This indicates that a single snapshot of a galaxy can give us a representative picture of its X-ray binary populations.

\subsection{X-ray emission as a SFR proxy for normal galaxies}
\label{sec:sfr_proxy}

\begin{figure}
\centering
{
\includegraphics[width=.6\textwidth]{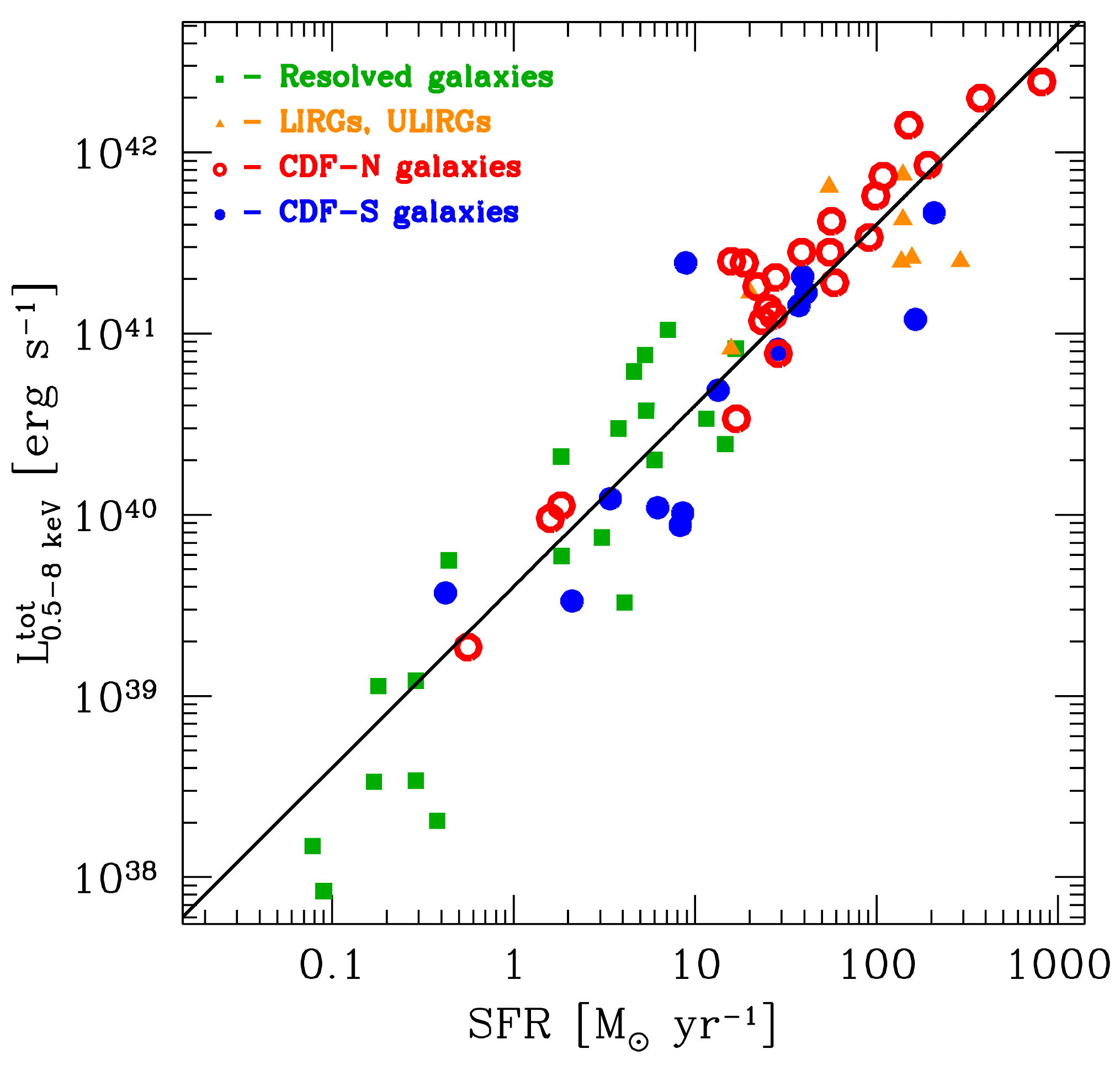}
}
\caption{The relation between SFR and total X-ray luminosity of normal galaxies. The solid line shows the linear scaling
relation $L_X=4\cdot 10^{39}\times {\rm SFR}$. Adapted from \cite{mineo14a}.} 
\label{fig:ltot_sfr}
\end{figure}

The promptness of HMXBs (Section \ref{sec:eta_hmxb}) makes them a potentially good tracer of the recent star formation activity in a galaxy \cite{stm78, ghosh2001}. Indeed, existence of  correlation of X-ray luminosity of star-forming galaxies with the classical SFR proxy -- FIR emission, has been noticed over three decades ago in the Einstein Observatory data (Section \ref{sec:intro}, refs. \cite{griffiths1990,david1992}). While there are other sources of X-ray emission in    star forming galaxies, such as ionized gas \cite[e.g.][]{mineo2012b}, in normal galaxies HMXBs dominate  and total X-ray emission correlates well with the SFR (Fig.\ref{fig:ltot_sfr}).

A multitude of various calibration methods are employed to estimate SFR in external galaxies, based on UV, H$_\alpha$, FIR or other wavelengths.  Any  SFR calibrator relies on certain assumptions concerning the environment in the galaxy which  lead to various uncertainties e.g. associated with the influence of dust, the escape fraction of photons, the shape of the IMF etc. In fact, many of the commonly used SFR indicators use the signatures of interaction of the ionizing emission from  massive stars with the interstellar medium, i.e. may suffer from the same systematic effects. A new  independent calibrator is therefore a useful addition to the  suite of SFR diagnostics employed by  modern astronomy.

A significant advantage of X-ray emission as a diagnostic tool  is its large penetrating power -- X-rays  are much less subject to attenuation by the neutral gas and dust than conventional SFR tracers. Galaxies are mostly transparent to X-rays above  $\sim 2$ keV, except for the densest parts of the most massive molecular clouds. The power law  spectra of X-ray binaries are also less susceptible to the effect of the cosmological redshift (K-correction).

The X-ray based SFR proxy, quite naturally, suffers from its own uncertainties and systematic effects. The most important of these is contamination by the activity of supermassive black hole, others are  related, for example, to the age and metallicity dependence of the X-ray binary populations. Stochastic effects and variability of the X-ray binaries, makes it less suitable for dwarf galaxies that are dominated by one or two HMXBs. Nonetheless, the X-ray emission of galaxies is a useful and promising proxy of the stellar populations and successfully  complements other conventional indicators, particularly in heavily obscured environments. There is a number of its successful applications to reconstruction of the cosmic star formation history \cite[e.g.][]{aird17, kurczynski2012, norman2004}

\subsection{Expectations from SRG/eROSITA all-sky survey}
\label{sec:erosita}

eROSITA X-ray telescope \cite{predehl2021} aboard SRG orbital observatory \cite{sunyaev2021} in the course of its on-going all-sky survey will detect of the order of $\sim 10^4$ normal galaxies out to the distance of a few hundred Mpc \cite{prokopenko2009, basu2020}. Although the moderate angular resolution of eROSITA  ($\sim 30"$ HPD averaged over the field of view) will not allow for detailed investigations of X-ray populations beyond Local Group, the might of the all sky coverage will make eROSITA data indispensable for detailed investigations of scaling relations and of the patterns of metallicity and age dependence. The main hurdle on this path will be identifying normal galaxies among over 3 million AGN and QSO dominating the eROSITA source catalog \citep{kolodzig2013}, and isolating the contribution of active nuclei of low luminosity. Significant source of contamination are also X-ray active stars.  To this end, multiwavelength data will play a critical role. \citet{galiullin2023} constructed the first SRG/eROSITA -- IRAS catalog of X-ray bright star-forming galaxies on the Eastern Galactic sky which currently (after 2 years of sky survey) includes of the order of $\sim 10^3$ star-forming galaxies. With this sample they studied dependence on the SFR and metallicity of the X-ray luminosity of starforming galaxies, separating it into contributions of X-ray binaries and hot inter-stellar medium (ISM) and connecting their finding with the Chandra results described earlier in this section.

\section{Spatial distribution of X-ray binaries in galaxies}

It was realized since the first X-ray observations of nearby normal  galaxies with the Einstein Observatory that the dominant emission above 2 keV -- dominated by XRBs - follows that of the stellar surface brightness \cite[see ][ and references therein]{fabbiano89}. With its sub-arcsecond angular resolution, Chandra has given us a unique opportunity to explore this connection, for both HMXB and LMXB populations. With Chandra observations  XRBs can be individually detected and associate  with different galaxy components in the optical band  \cite[see ][ and references therein]{fabbiano}. 

For the HMXB population, two results are particularly notable: 1) The definitive association of ULXs with star-forming regions, (e.g., in the Antennae \citep{fabbiano2001}, the Cartwheel galaxy \citep{wt2004}, nearby colliding galaxy pair NGC 2207/IC 2163 \citep{mineo2013}), which supports the suggestion that most ULXs may be super-Eddington accreting HMXBs \citep{king2001, soria2009, gm2015}. 2) The lack of HMXBs in region of very recent intense star formation, which is consistent with the evolutionary time for a massive binary to reach the HMXB stage. This effect was observed in M51 \citep{shtykovskiy2007a}  and the Magellanic Clouds \cite{shtykovskiy2005a,Antoniou2016}.

The study of the spatial distribution of LMXBs in elliptical galaxies has been motivated by the desire to understand the prevalent formation mechanism for these systems: from the evolution of field binaries, or from dynamical formation in Globular Clusters (GC). In the first case, the radial distribution of LMXB may follow closely that of the stellar light, while in the second it may be more extended. While early studies were inconclusive \citep[see review in][] {fabbiano}, more recently results show a possible excess of LMXB at larger radii than expected from the distribution of the optical surface brightness of the galaxies \citep{zhang13, mineo14b}. This line of investigation is further discussed in Section \ref{sec:lmxb_spatial_distr2}.

\section{Primordial and dynamically formed LMXBs}
\label{sec:dynamical}

\subsection{LMXB formation channels}

LMXB populations
are expected to form through two basic pathways: (1) Roche-lobe overflow of
low-mass ($\simlt$2--3~$M_\odot$) donor stars onto compact-object companions in
isolated binary systems that form in situ within galactic fields; and (2)
dynamical interactions (e.g., tidal capture  and multibody exchange with
constituent stars in primordial binaries) in high stellar density environments
like globular clusters \citep[GCs,][]{clark1975,fabian1975,hills1976} and
near the centers of galaxies \citep[e.g.,][]{voss07b,zhang11}. The relative roles of these two channels have been debated. Given the very high formation efficiencies of GC LMXBs (a factor
of $\approx$50--100 times higher per stellar mass than that of the field), it has been speculated that field LMXB populations may initially
form dynamically within GCs, and subsequently ``seed'' galactic fields
through ejection or GC dissolution \citep[e.g.,][]{grindlay84, kremer18}
{\it Chandra} observations have shown strong evidence that both formation
channels are important, with the normalization of the LMXB XLFs being primarily
dependent on the integrated stellar mass, with an additional significant
dependence on GC specific frequency \citep[number of GCs per galaxy stellar mass;
see, e.g.,][]{lmxb,irwin05,juett05,humphrey08,kim09,boroson11,zhang11,lehmer20}.

The combination of \chandra\ and {\it Hubble Space Telescope} surveys of
relatively nearby ($D \simlt 30$~Mpc) early-type galaxies have allowed for
disentanglement of field and GC LMXB populations through the direct
multiwavelength classification of \xray\ point sources \citep[e.g.,][]{kim09,voss09,paolillo11,luo13,lehmer14,lehmer20,mineo14b,peacock16,peacock17,dage19}.
Studies of LMXBs directly associated with GCs have revealed that their formation efficacy
depends on both local stellar
interaction rates and metallicity \citep[see, e.g.,][]{pooley03,sivakoff07,cheng18}.  The most apparent trends appear in observed
GC colors, with metal rich, red, GCs hosting a factor of $\approx$3 times more
LMXBs compared to metal poor, blue GCs \citep[e.g.,][]{kim13}; however, no
significant variation in the GC LMXB XLF slope is observed as a function of GC
metallicity.

\begin{figure*}
\centerline{
\includegraphics[width=0.7\textwidth]{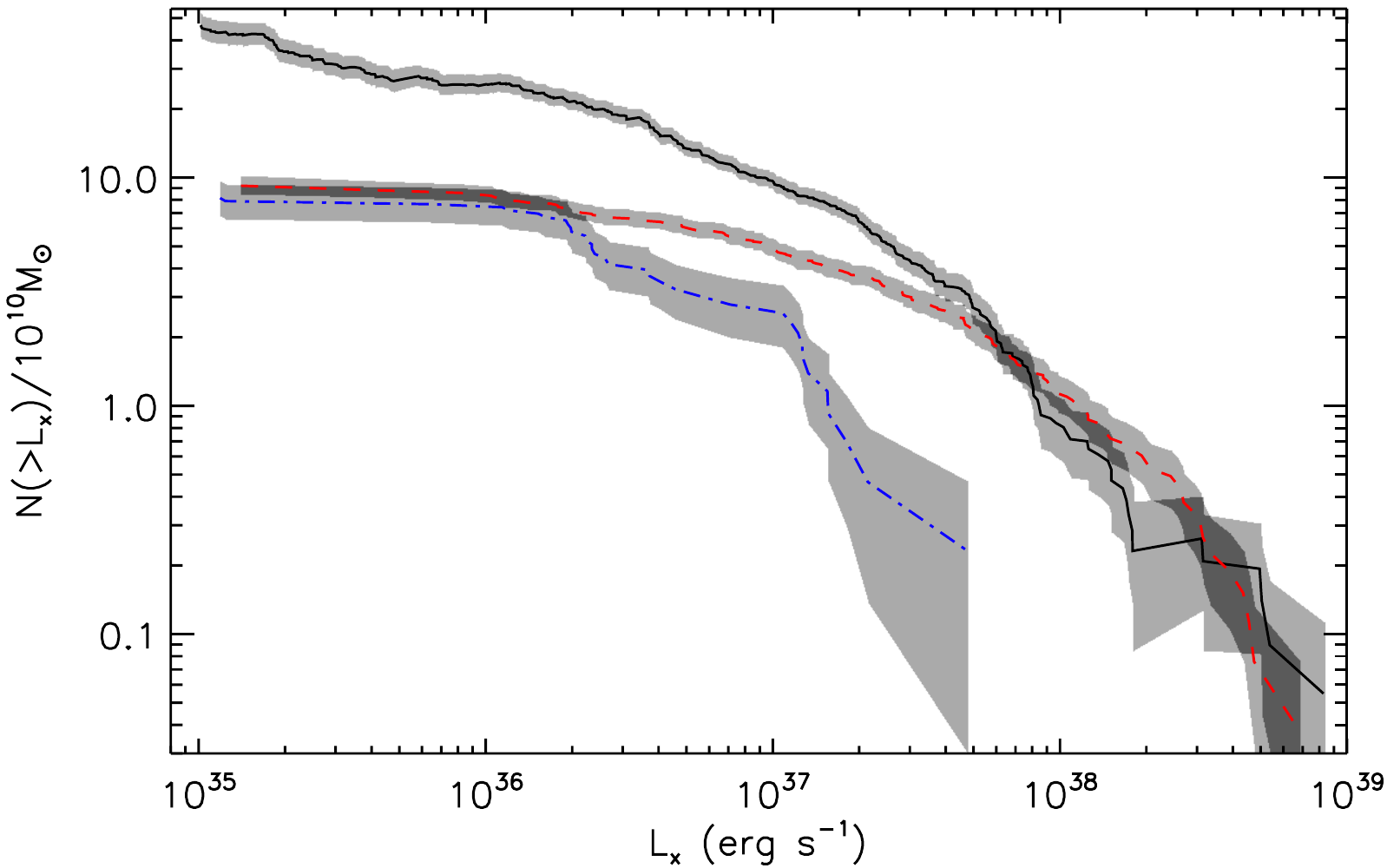}
}
\caption{The stacked XLFs of LMXBs in different environments: field and GC sources in early type galaxies and nucleus of M31. The contribution of CXB sources was subtracted and the incompleteness correction was applied. The field XLF (solid) is normalized to the stellar mass of $10^{10}~M_\odot$. The normalizations of GC (dashed) and M31 nucleus (dash-dotted) XLFs are arbitrary. The shaded areas around the curves show $1\sigma$ statistical uncertainty.
Adapted from \citep{zhang11}.
}
\label{fig:lmxb_xlf_gc_field}
\end{figure*}

\subsection{Clues from  luminosity functions}

\begin{figure*}
\begin{minipage}{0.38\textwidth}
\includegraphics[width=\textwidth]{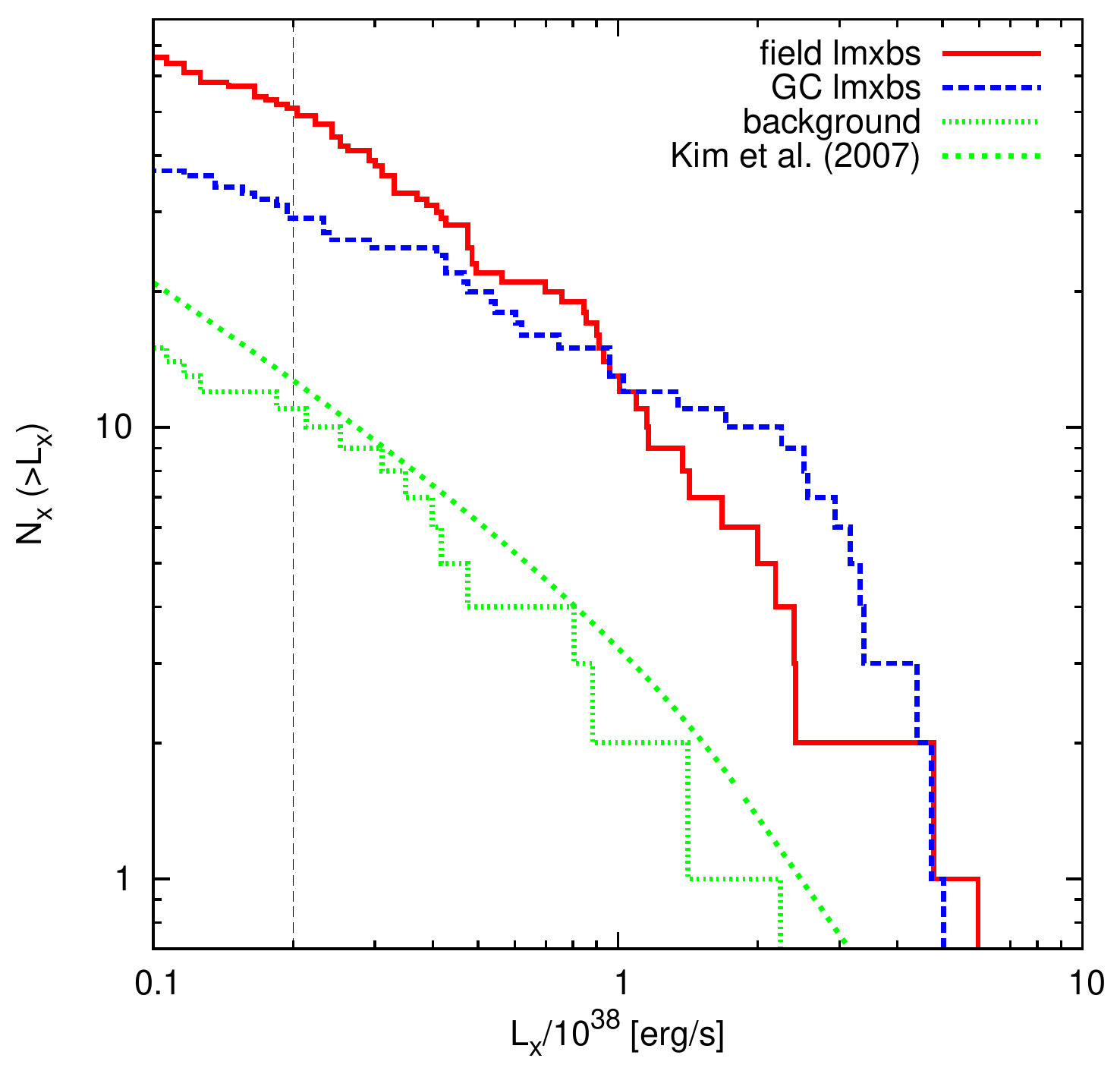}
\vspace{1mm}
\end{minipage}
\hspace{2mm}
\begin{minipage}{0.58\textwidth}
\includegraphics[width=\textwidth]{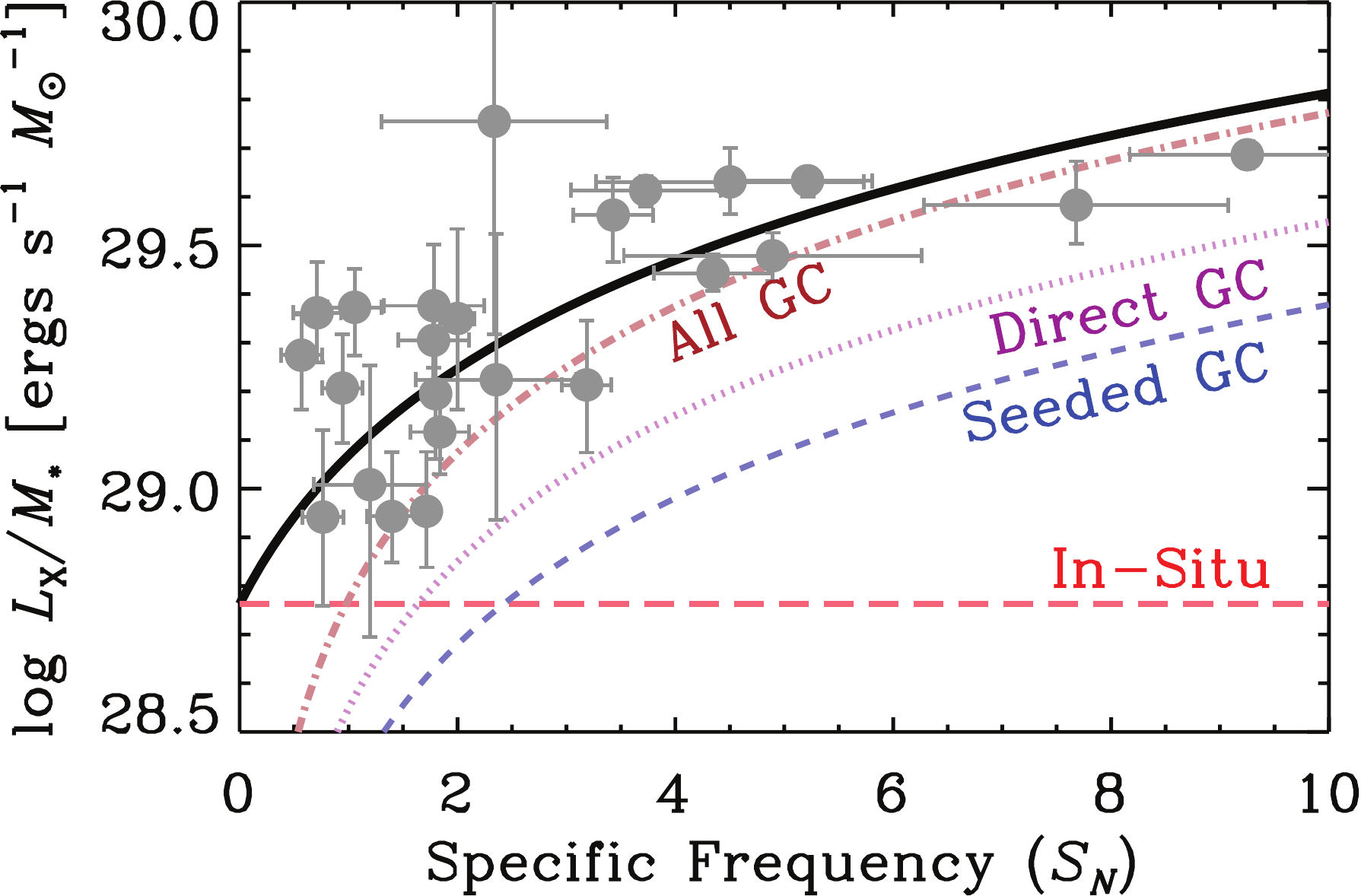}
\end{minipage}
\caption{ ({\it Left}) Cumulative field ({\it red solid}) and GC ({\it blue dashed}) LMXB luminosity functions for NGC~4594 from \citep{peacock14}, illustrating the flatter XLF of GC LMXBs relative to field LMXBs. ({\it Right}) X-ray luminosity per stellar mass versus GC specific frequency, $S_N$, for a sample of 24 nearby elliptical galaxies \citep{lehmer20}.  Contributions are shown for LMXBs that are formed ``in situ'' within the galactic fields ({\it red long-dashed}), those that are observed in galactic fields, but suspected of being formed in, and ejected from GCs ({\it blue short-dashed}), and those that are directly coincident with GCs ({\it purple dotted}).
}
\label{fig:gcvfield}
\end{figure*}

Variations in the shapes of the LMXB XLFs between field and GC environments have  provided valuable insights into the nature and evolution of the LMXB population.   
Studies of the field and GC LMXB XLFs revealed that the
shapes of the XLFs  differ significantly, with the GC LMXB XLF appearing flatter than that of the field (Fig.\ref{fig:lmxb_xlf_gc_field}, \ref{fig:gcvfield})  \citep{zhang11, peacock14, peacock16}.  Furthermore, reports of an age dependence, in which the
number of field LMXBs per galaxy stellar mass declines with increasing light-weighted stellar age \citep[e.g.,][]{kim10,lehmer14},
suggested that the field LMXB XLF appeared to evolve with stellar age, a result
expected from XRB population synthesis models \citep[e.g.,][]{fragos09,fragos13}
and indirectly observed in deep \chandra\ surveys as an increase of the $L_{\rm
X}$(LMXB)/$M_\star$ scaling relation with increasing redshift \citep[e.g.,][]{lehmer07,lehmer16,aird17}.  However, these studies suffered from small
number statistics and large uncertainties on light-weighted stellar ages.

A more recent study by \citep{lehmer20} culled 24 nearby early-type galaxies
with \chandra, {\it Hubble}, and additional multiwavelength observations with
the aim of investigating the nature of the field LMXB XLF.  This work showed
that early-type galaxies contain stellar-mass weighted ages that span only a
narrow range where in-situ LMXB XLFs are not expected to vary.  Global modeling
of the field LMXB XLFs required scaling from both stellar mass and GC specific
frequency with high confidence.  Furthermore, the shape of the GC-related
field LMXB XLF component was shown to be consistent with the XLF of LMXBs
directly coincident with GCs, suggesting that some seeding of the field LMXB
population from GCs is likely and significant in massive early-type galaxies.  The right panel of Figure~\ref{fig:gcvfield} shows the level of contributions expected to the integrated $L_{\rm X}[{\rm LMXB}]/M_\star$ as a function of $S_N$ for in-situ LMXBs, GC seeded LMXB observed in galactic fields, and LMXBs directly coincident with GCs.

\subsection{Clues from the spatial distributions}
\label{sec:lmxb_spatial_distr2}

\begin{figure*}
\hbox{
\begin{minipage}[b][0.6\textwidth][s]{0.48\textwidth}
\vspace{0.4cm}
\includegraphics[width=\textwidth]{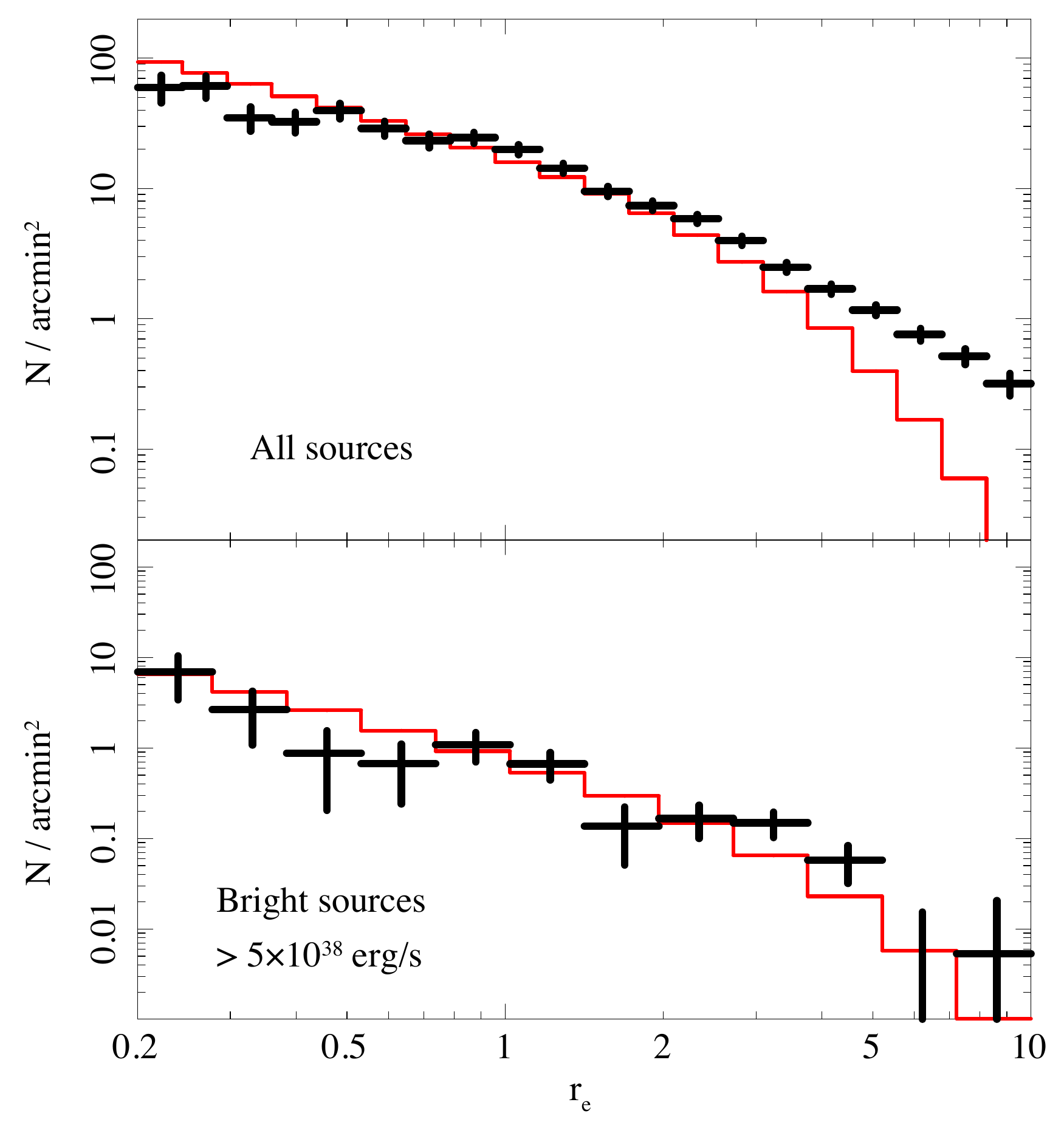}
\end{minipage}
\hspace{0.2cm}
\begin{minipage}[b][0.6\textwidth][s]{0.5\textwidth}
\vbox{
\includegraphics[width=0.9\textwidth]{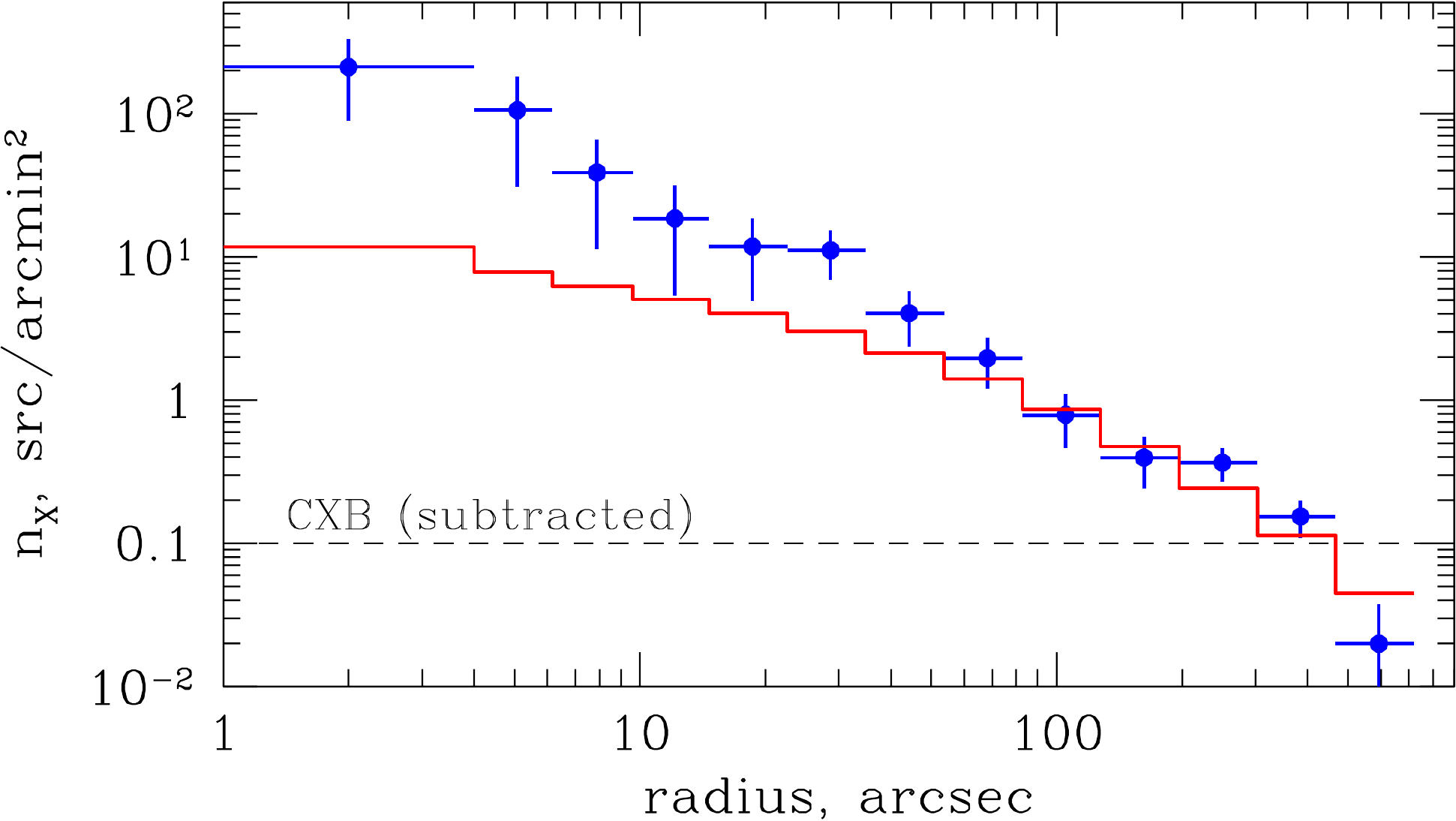}
\hbox{
\hspace{0.2cm}
\includegraphics[width=0.85\textwidth]{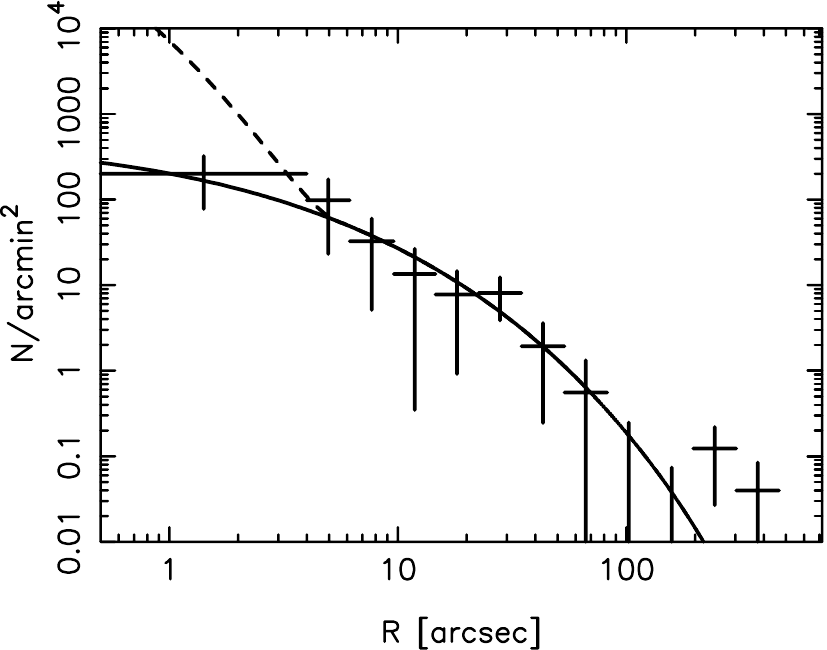}
}
}
\end{minipage}
}
\caption{{\em Left:} Stacked radial source density profiles of LMXBs in nearby early type galaxies, segregated by the source luminosity. Contribution of CXB sources subtracted. Solid histograms show predicted 
distribution based on the k$_s$-band light. The x-axis shows radial distance in units of the effective radius.
Adapted from \citep{zhang13}.
{\em Right:} Dynamical formation of LMXB in the nucleus of M31. The top panel shows radial distribution of LMXBs in the nucleus of M31, excluding globular cluster sources. The CXB level is subtracted (shown by the dashed line). 
The solid histogram shows the projected distribution of the stellar mass. The normalization of the latter is from the best fit to the data outside 1 arcmin. The bottom panel shows  distribution of the ``surplus'' X-ray sources, computed as a difference between the data and the stellar mass model shown in the top panel.
The solid line shows the projected  $\rho_*^2$
distribution, computed from the  same mass model of the M31 bulge. Adapted from \citep{voss07b}
}
\label{fig:lmxb_spatial}
\end{figure*}

Evidence supporting at least some contribution of GC seeding has also been observed in the spatial
distributions of LMXBs in early-type galaxies.  \citet{zhang13} showed that
the radial distributions of LMXBs in 20 early-type galaxies mainly followed the
stellar light profiles, but contained an excess of low-luminosity ($\simlt5
\times 10^{38}$~ergs~s$^{-1}$) \xray\ sources out to $\approx$10 effective
radii (Fig. \ref{fig:lmxb_spatial}).  Such an excess is consistent with being associated with blue (metal poor) GCs,
which follow broader profiles relative to stellar light, however it is also observed in galaxies with low GC content -- the extended LMXB halos must be a combined result of GC seeding and neutron-star LMXBs being kicked out of the main bodies of their host galaxies by asymmetric supernova
explosions \citep{zhang13}.

High stellar densities where stellar interactions become important are also reached in the galactic nuclei -- similar to globular clusters they may become the  site of the dynamical formation of LMXBs. \citet{voss07a, voss07b} found a significant increase of the specific frequency of X-ray sources in the nucleus of M31 (Fig.\ref{fig:lmxb_spatial}). The large volume of the bulge and correspondingly large number of dynamically formed LMXBs (about $\sim 20$ with $L_X>10^{36}$ erg/s) permitted them to directly probe the density profile of dynamically formed binaries which  was shown to follow $\rho_*^2$ law (Fig.\ref{fig:lmxb_spatial}, right bottom panel). As galactic bulges have $\sim 1-1.5$ orders of magnitude larger velocity dispersion than globular clusters, the details of stellar interactions and composition of dynamically formed binary populations are different in bulges and  clusters \citep{voss07b}.

Detailed investigations of the spatial distributions of
GC populations in the elliptical galaxies NGC~4261, NGC~4649, and NGC~4278 show
anisotropies that are consistent with anisotropies observed in the distributions of both GC and field LMXBs \citep{zezas03,dabrusco14a,dabrusco14b}. The GC anisotropies show arc and streamer-like morphologies,
indicative of past dwarf-galaxy mergers.  The observed anisotropies in the field LMXB
distributions could be due to GC LMXB ejection or could be relics of past star-formation in the merging systems.

A particularly good study case is that of NGC 4649, a giant E in the Virgo cluster. This study was made possible by the coordinated complete deep coverage of both GC and LMXB populations with HST and Chandra surveys, which resulted in reliable identifications of X-ray sources with GC counterparts \citep{strader2012,luo13}. Using these rich data sets, Mineo et al. \citep{mineo14b} concluded that the evolution of field binaries and the dynamical formation in GCs are both likely to occur: LMXBs spatially coincident with GCs follow the same radial distribution as the overall distributions of red and blue GCs, instead those with no GC counterpart are radially distributed like the stellar surface brightness, except perhaps at larger radii. Interestingly, in the 2-dimensional distributions of LMXBs and GCs on the plane of the sky in NGC 4649  there are arc-like distributions of GCs associated with similar over-densities of LMXBs \citep{dabrusco14a,dabrusco14b}. However, a significant localized over-density of field LMXBs is found to the south of the GC arc, suggesting that these LMXBs may be somewhat connected with the arc of GCs. These sources occur at relatively large galactocentric radii and could contribute to the excess of field LMXB reported in \citet{mineo14b}.

\section{Ultra-luminous X-ray sources.}

An unusual class of compact sources -- ultraluminous X-ray sources (ULXs), was discovered in the first survey of nearby galaxies with the Einstein Observatory in the late 1970s
(see reviews \citep{fabbiano89, kaaret2017}).   What characterizes these sources is their X-ray luminosity  $L_X>10^{39}$ erg/s, in excess of that expected from the Eddington limit for an object of ~1-5 solar mass. This was at time the range of known masses  for the compact accretor in  Milky Way XRBs ( neutron stars and stellar mass black holes).  This discovery led to the speculation that ULXs could represent an entirely new class of astrophysical objects, Intermediate Mass Black Holes (IMBH), with masses ~$100 - 10^4 $ solar mass, bridging the gap between the stellar black holes  and the supermassive $10^8$ solar mass black holes at the nuclei of galaxies \citep[e.g.][]{colbert99}. 

However, a number of alternative models were advanced
as well -- from collimated  radiation \citep{koerding2002} to  $\sim$stellar mass black holes, representing the high mass tail of the 
standard stellar evolution sequence and accreting in the near- or slightly super-Eddington regime \citep{king2001,grimm03,gm2015}. Also, a critical review of the observational properties of ULXs demonstrated that their association to IMBH, while possible, was not proven beyond reasonable doubt \citep{fabbiano05}.

As discussed below, Chandra observations have shown a prevalent (but not unique) association of ULXs with regions of intense star formation in galaxies, supporting the hypothesis of a stellar nature for these objects. Recent gravitational wave observations have also demonstrated that stellar BHs can have much higher masses than previously thought (up to ~100 solar masses \citep{barrera22} \footnote{see also https://www.ligo.caltech.edu/image/ligo20171016a}). More recently, the NuSTAR discovery of several pulsating ULXs \citep{bachetti14, walton18}, has changed dramatically the landscape of theoretical models, confirming the old theoretical prediction than the accretion column on the magnetic pole of a neutron star is less subject to the Eddington limit constrain \citep{basko1975, basko1976,king2001}.

The present view on the nature of ULXs is more nuanced. It is quite possible that these luminous sources do not represent a unique astrophysical class of object. They may include super-Eddington XRBs (either NS of BH binaries), and also IMBH, especially in the case of very luminous ULXs at the outer radii of their associated galaxy \citep[e.g.][]{kim15}.

\subsection{Association with star-formation}
The XLFs of compact X-ray sources in nearby early and late type galaxies (Fig. \ref{fig:xlf}) make it quite obvious that luminous X-ray sources with $L_X\simgt 10^{39}$ erg/s are present in significant numbers only in star-forming galaxies where XLF extends into the range of luminosities attributed to  ULXs, to $L_X\sim 10^{40}$ erg/s. Detailed studies of  X-ray binary populations in  individual galaxies showed that ULXs are preferentially located in or near star-forming regions \citep[e.g.][]{colbert04, mineo2013, zezas02, zezas07}. 
Furthermore, comparison between the location of the ULXs and their nearest star-clusters or star-forming regions set stringent constraints on the strength of surpernova kick velocities \citep{zezas02,kaaret2004}. 
 
These results have been confirmed by more recent studies of ULXs in large samples of galaxies observed with \chandra{} \cite{kovlakas2020,swartz2011}, which showed that ULXs are predominantly located within late-type galaxies with high sSFR. A particularly strong trend is the prevalence of ULXs in galaxies with low-metallicity  \citep{mapelli2009,brorby16,kovlakas2020,prestwich2013}. This trend is attributed to the weaker stellar winds of lower metallicity stars, resulting in tighter orbits and hence larger fraction of systems undergoing very efficient mass transfer \citep[e.g.][]{Linden2010,fragos13}.   

In contrast, searches for ULXs in early-type galaxies showed that they are rather scarce and are found predominantly in young ellipticals (Fig.\ref{fig:xlf2}). Zhang, Gilfanov and Bogdan \citep{zhang12} found 24 sources with $L_X>10^{39}$ erg/s within $D_{25}$ in a sample of 20 early type galaxies with measured stellar age, the expected number of background AGN being equal to 11.8. The luminous sources are mostly associated with younger galaxies -- 17 and 7 in galaxies younger and older than 6 Gyrs, with the CXB sources expectation of 5.8 and 6 respectively. The specific frequencies of luminous sources are $8.8\pm 3.2$ sources per $10^{12}$ M$_\odot$ in young galaxies with the 90\% upper limit of 2.9 sources  per $10^{12}$ M$_\odot$ in galaxies older than 6 Gyrs \citep{zhang12}.

Similarly,  Kovlakas and collaborators \citep{kovlakas2020} found a small number of ULXs in early-type galaxies which appears to correlate non-linearly with the stellar mass of their host galaxy. They interpreted this behaviour in the context of variations in the star-formation history of the galaxies, in agreement with ULX population synthesis models \cite{kovlakas2020}.

\subsection{Main conclusions from optical studies}

The association of ULXs with intensely star-forming galaxies provided the first indications that they are a sub-class of HMXBs. However, systematic studies of their environment showed that ULXs are often located inside bubbles of ionized gas as witnessed for strong He$_{\rm{II}}$, or OIII$_{\rm{III}}$ lines \cite[e.g.][]{kaaret2009}. While many of these bubbles are shock-excited by outflows from the ULX, the presence of strong excitation lines (e.g. He$_{\rm{II}}$, Ne$_{\rm{V}}$) in some of them clearly indicates photoinization by a hard ionizing source. 
In this case they can provide a direct measure of the soft X-ray luminosity of the ULX and therefore an stringent constraints on any beaming. The circular shape of many of these bubbles and the high inferred ionising  luminosity 
(e.g. $>10^{40}\rm{erg\,s^{-1}}$ in the case of HoII-X1 \citep{Berghea2010}) suggests mild beaming.

Searches for optical counterparts to ULXs have been successful for $\sim20$ sources \cite{Tao2011,Gladstone2013}. These tend to have V-band luminosities similar to early-type star supergiants and blue colours. However, their optical SEDs are not consistent with stellar spectra, indicating that they are dominated by the accretion disk component  \cite{kaaret2017}.

\begin{figure} 
\centering
\includegraphics[width=.6\textwidth]{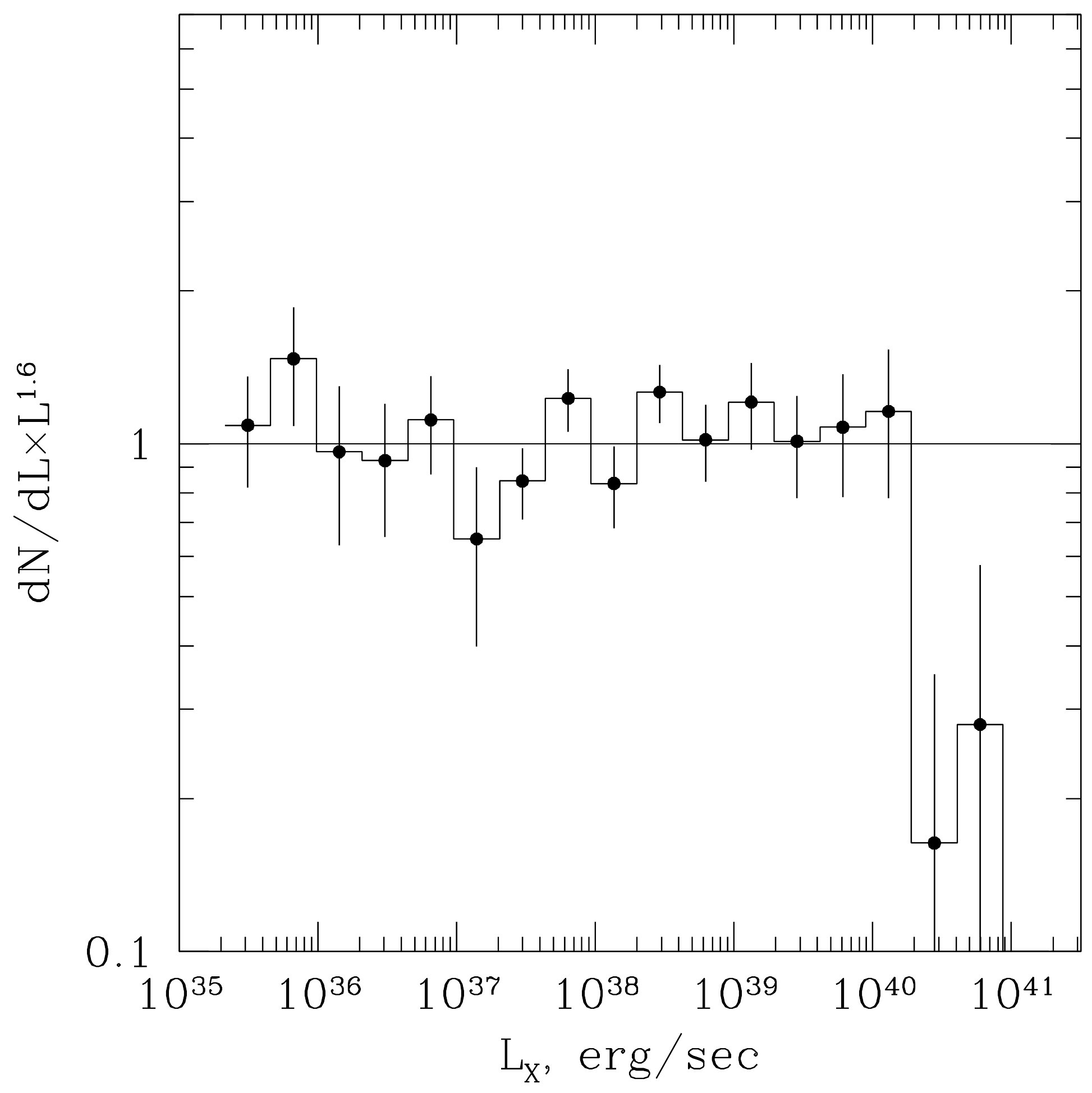}
\caption{Detailed shape of the X-ray luminosity function of compact X-ray sources in star-forming galaxies. The figure shows the ratio of the X-ray luminosity function from Fig.\ref{fig:xlf} to a power law with slope of 1.6. Based on results of \cite{mineo12}. }
\label{fig:xlf_rat}
\end{figure}

\subsection{Inferences from the shape of the HMXB luminosity function}

As it has been well known,  the maximum mass of a black holes  produced in the course of standard stellar evolution at solar abundance of 
elements is limited to $\approx 10-20$ M$_\odot$, whereas formation of more massive black holes with mass exceeding $\sim 100$ is only 
possible at virtually zero abundance of metals \citep{zhang08}. It is not quite clear yet, whether this conventional  picture contradicts the LIGO detections of $\sim 100$ M$_\odot$ black holes in coalescing binaries \citep{barrera22}. It is possible in principle that the most luminous sources are accreting 
IMBHs -- descendants of Pop III stars, which  acquired a massive companion in star-forming regions. Obviously, the abundance of 
such systems should be significantly smaller than abundance of normal high-mass X-ray binaries formed in the course of standard stellar 
evolution. Therefore there must be a break in the luminosity function at the transition between "normal" X-ray binaries and these objects. 
However, observations show that the luminosity distribution of compact  X-ray sources in star forming galaxies smoothly extends up to the 
luminosities of $\log L_{\rm X}\sim 40-40.5$, without any significant features or slope changes. In particular, unlike the LMXB LXF, it 
does not have any significant features at the luminosities corresponding to the Eddington limit of a neutron star ($\log L_{\rm X}\sim 38.3$)
or of a black hole ($\log L_{\rm X}\sim 39-39.5$). On the other hand, it breaks at the luminosity $\log L_{\rm X}\approx 40.0-40.5$ 
(Fig.\ref{fig:xlf_rat}), corresponding to the Eddington luminosity of a $\sim 100 $ M$_\odot$ object. 
Because of such a smooth shape of their XLF,  it appears most likely that systems with luminosity  $\log L_{\rm X}\le 40-40.5$ are ``normal'' 
X-ray binaries formed in the course of standard stellar evolution and represent the tail of the distribution of black hole masses and mass 
accretion rates. We note here that luminosities exceeding the Eddington limit by several times  are possible in the standard accretion model 
\citep{ss73, grimm03}. The break in the HMXB XLF observed at $\log L_{\rm X}\sim 40.5$ (Fig. \ref{fig:xlf},\ref{fig:xlf_rat}) may indicate the 
transition to a different population of X-ray sources. The few known sources with luminosities exceeding this value may indeed be IMBHs -- result of the evolution of Pop III stars.  

\subsection{Possible nature and implications for accretion physics}

The nature of ULXs has been a matter of debate since their discovery in the early 1980s, with potential models including  super-luminous supernova remnants, accreting IMBHs, highly accreting X-ray binaries formed through dynamical capture or secular evolution \citep[see e.g.][for a critical summary of these models]{zezas02,fabbiano,kaaret2017}.  Mounting evidence based on the their X-ray luminosity distribution, association with young stellar populations, scaling with SFR and metallicity, and their multi-wavelength counterparts or surrounding nebulae, suggests that the vast majority of ULXs are HMXBs undergoing a rapid mass-transfer episode. In this respect they are the upper end of the X-ray luminosity range of HMXBs. In fact, detailed modeling of the mass-transfer sequences of HMXBs \citep[e.g.][]{Rappaport2005} indicates that systems with massive donors ($\simgt 10 {\rm M}_\odot$) can undergo brief phases ($\sim 10^{2}-10^{4}$\,yr) of mildly super-Eddington accretion resulting in X-ray luminosities even in excess of $\sim 10^{40}$~ergs~s$^{-1}$. These episodes take place at the thermal-timescale of the donor star.
These results are supported by a growing volume of X-ray binary population-synthesis models which show that indeed HMXBs can experience brief phases of super-Eddington mass transfer which in some cases can reach accretion rates well in excess of $10^{3}$\,$\rm{\dot{M}_{Edd}}$;  the fraction of these super-Eddington systems increases dramatically in lower metallicities \citep[e.g.][]{Linden2010,Wiktorowicz2017,Marchant2017}.
 
Further evidence for supercritical accretion comes from the discovery of pulsar-ULXs which gives a direct constraint on the compact object mass \citep[e.g.][]{bachetti14}
A natural outcome of supercritical accretion is the formation of the so-called slim or thick accretion disks. The slim disks are expected to form at accretion rates exceeding $\sim\rm{M_{Edd}}$, and they have larger height (thickness) than the standard thin disks \citep{ss73,slim_disk96}, resulting in non-linear scaling of the emerging X-ray luminosity with accretion rate and different spectral shape.  In higher accretion rates the radiation pressure further increases its thickness resulting in the formation of a funnel in its inner part \citep{slim_disk96}. A natural outcome of this effect is mild beaming of the X-ray emission \citep[e.g.][]{king2002}. 
Recent general relativistic, radiation magneto-hydrody namical simulations of supercritical mass accretion models for black holes confirmed this picture \citep[e.g.][]{Sadowski2016}.  
 
The differences in the structure of the accretion disk in ULXs with respect to lower-luminosity X-ray binaries becomes evident in their X-ray spectra,  which often show a curvature above $\sim2$ keV which can be described by a break at $\sim7-10$ keV  \citep[][and references therein]{poutanen2007, kaaret2017}. Evidence of reflection of the pulsar emission from the walls of the accretion funnel has been also found \citep{bykov2022}.

\section{Population synthesis results}
\label{sec:pop_synthesis}

\subsection{Relevant results from binary evolution}

X-ray binary evolution models have provided important insights in the formation timescales of the different types of X-ray binaries. These are determined by a combination of the nuclear  timescales of the donor stars and the time required for the onset of mass transfer from the donor star to the compact object. 
The mass transfer can take place in two main ways: (a) capture of material expelled from the donor star in the form of a stellar wind, and (b) Roche-lobe overflow.  The stellar winds are mostly relevant in the case of X-ray binaries with early-type donors (O,B stars, supergiants, or Wolf-Rayett stars), since lower-mass stars  have very weak stellar winds that cannot produce observable X-ray emission. A special case of wind-fed X-ray binaries are the Be X-ray binaries (Be-XRBs) where the accreted material originates in an equatorial outflow (decretion disk) from the donor Oe or Be star \citep[e.g.][]{Reig2011}. On the other hand, the onset of Roche-lobe overflow (RLOF) requires that the radius of the donor star becomes larger than its Roche lobe. This usually happens either as a result of the increase of the stellar radius as the star evolves off the main sequence, and/or as a result of the shrinkage of the orbital radius of the system (e.g. due to tidal evolution, magnetic breaking, common-envelope evolution etc).

Since the donor stars have lower mass than the primary star producing the compact objects, their nuclear evolution timescales are longer.  High-mass X-ray binaries appear between a few Myr and $\sim100$\,Myr from the formation of the binary stellar system.  The low limit is driven by the time required for the most massive star of the system to produce a compact object. The  upper range reflects the upper range of the timescale needed for the donor star (which in the case of HMXBs is of O or B spectral types) to evolve and initiate the mass transfer. 
In the case of Low Mass X-ray binaries, their formation timescales are much longer, from a few hundred Myr up to several Gyr. This is because because of the longer evolutionary timescales of lower-mass donor stars, and the long timescale required for the shrinkage of the orbital radius.
 
Binary stars are formed with a wide range of mass ratios, orbital separations, and eccentricity. Only a small fraction of these systems will eventually become X-ray binaries (Section \ref{sec:xrb_freqs}). Even before the formation of a compact object as a result of the nuclear evolution of the more massive star, the two stars may interact, exchanging mass. The supernova explosion may have a dramatic effect in the evolution of the system by imparting a kick onto the resulting compact object. The result of the kick is to increase the orbital separation and/or the eccentricity of the orbit. In the most extreme case it may disrupt the binary system. However, as a result of the nuclear evolution of the secondary, mass transfer either through a stellar wind or RLOF may initiate, resulting in an X-ray binary. 
If the mass of the donor star is lower than $\sim10 {\rm M_\odot}$  the initiation of mass transfer also requires shrinkage of the orbit. This can take place through a variety of mechanisms: magnetic breaking, tidal interaction, emission of gravitational radiation. A particularly effective mechanism is common envelope evolution \citep[e.g.][]{Taam2000, Ivanova2020}, which however, in many cases may lead to the merging of the two systems.

\subsection{Summary of population synthesis models and their results}

X-ray binary population synthesis models are an invaluable tool for understanding the X-ray binary populations, their connection with fundamental parameters of the host galaxy and for constraining uncertain parameters of the theory through comparisons with observations.
They calculate the populations of X-ray binaries at a given time of the evolution of a stellar population by combining distributions of the initial parameters of the binary systems together with prescriptions for the evolution of the stars in the binary systems and its orbital parameters \citep[e.g.][]{belczynski08,COMPASS,fragos13}. 
As a result, one then can model the X-ray emission of different X-ray binary populations as a function of the star-formation history of their parent stellar populations. 

Figure \ref{fig:lx_age} shows the dependence of the integrated X-ray luminosity of a stellar population as a function of its age and metallicity. It is clear that X-ray binary populations associated with younger and lower metallicity stellar populations tend to have higher X-ray luminosities. The metallicity dependence is particularly important for understanding the cosmological evolution of X-ray binary populations and their potential role in the early Universe (Section \ref{sec:evolution}).

The first attempts to model the X-ray binary populations observed with \chandra{} using population synthesis showed that despite the large number of parameters in these models, their results are relatively robust \citep[e.g.][]{belczynski08,COMPASS,fragos13}.  Nonetheless, comparison of the X-ray luminosity functions of X-ray binaries obtained in \chandra{}  led to useful constraints on parameters such as the strength of the stellar winds, the common-envelope ejection efficiency, and the  mass-ratio of the stars in the binary system at the zero-age main sequence \cite{Tzanavaris2013}. Similar constraints can be set by looking at the integrated X-ray emission of  X-ray binaries in unresolved galaxies \citep[e.g.][]{fragos13,lehmer16}. 
Furthermore, comparison of the measured age-evolution of the formation rate of X-ray binaries with predictions from X-ray binary population synthesis models showed (Fig. \ref{fig:eta_hmxb}) that they reproduce well the increased populations at ages $\sim10-50$ Myr and the decline of their integrated X-ray luminosities at systems older than $\sim5$ Gyr, as well as, their metallicity dependence \citep{shtykovskiy2007b,Antoniou2010,Antoniou2016, Antoniou2019,lehmer21}.

In the case of ULXs, population synthesis models showed that their increased rates at low metallicities is driven by the reduced effect of stellar winds in removing mass and angular momentum from the system and hence resulting in tighter orbits and a larger fraction of systems undergoing RLOF mass transfer \citep[e.g.][]{Linden2010}. Furthermore, detailed modeling of individual ULXs (and especially pulsar-ULXs) revealed additional formation channels for these rare but very luminous systems \citep[e.g.][]{Misra2020,Abdusalam2020}.

\subsection{How frequent are X-ray binaries?}

\label{sec:xrb_freqs}

With the knowledge of the relation between the number of HMXBs and SFR, we can estimate the fraction of compact objects that become HMXBs \citep{mineo12}.
According to the HMXB XLF and scaling relation (Section \ref{sec:xrb_scaling}), the number of HMXBs with luminosity higher than $10^{35}$ erg/s is:
\begin{equation}
N_{\rm{HMXB}}(>10^{35} \rm{erg}\,\rm{s}^{-1}) \approx 135 \times \rm{SFR} 
\label{eq:Nhmxb_lowL}
\end{equation}
On the other hand, the number of HMXBs  can be expressed via the birth rate of compact objects $\dot{N}_{co}$ :
\begin{equation}
N_{\rmn{HMXB}}\sim \dot{N}_{co}\, \sum_k f_{\rmn{X},k}\, \tau_{\rmn{X},k} 
\sim \dot{N}_{co} \, f_{\rmn{X}} \,  \bar{\tau}_{\rmn{X}}
\label{eq:Nhmxb_anytime}
\end{equation}
The $\dot{N}_{co}$ approximately equals to the birth rate of massive stars $\dot{N}_{co}\approx\dot{N}_{\star}(M>8\,M_{\odot})\approx 7.4 \cdot 10^{-3} \times \rmn{SFR}$.\footnote{use of the Salpteter IMF is explained in \cite{mineo12}} The $f_{\rmn{X}}=\sum_k f_{\rmn{X},k}$  is the  fraction of  compact objects  which become X-ray active in HMXBs and  $\bar{\tau}_{\rmn{X}}$ is the average X-ray life time of such objects.  As discussed in \cite{mineo12}, the low and moderate luminosity sources are dominated by Be/X systems, therefore $\bar{\tau}_{\rmn{X}}\sim 0.1$ Myr (cf.Fig.\ref{fig:eta_hmxb}) \citep{verbunt}.  
Combining equations \ref{eq:Nhmxb_lowL} and \ref{eq:Nhmxb_anytime}, we obtain:
\begin{equation}
f_{\rmn{X}} \sim 0.18 \times \bigg(\frac{\bar{\tau}_X}{0.1\,\rmn{Myr}}\bigg)^{-1}
\label{eq:fraction_ns_bh_hmxb}
\end{equation}

Thus, we arrived to a remarkable conclusion that a large fraction, of the order of $\simgt$ten per cent of all black holes and neutron stars once in their lifetime were X-ray sources with $L_\rmn{X}> 10^{35}$ erg/s, powered by accretion from a massive donor star in a high-mass X-ray binary. 

Similarly, given the scaling relation for ULXs $N_{\rm ULX}(>10^{39} {\rm erg/s}) \approx 0.48 \times {\rm SFR}$ one can show that 
\begin{equation}
f_{\rm ULX} \sim 3.5\cdot 10^{-2} \times \bigg(\frac{\bar{\tau}_{\rm ULX}}{10^{4} \,\rm{yr}}\bigg)^{-1} 
\label{eq:fx_ulx}
\end{equation}
i.e. a few per cent of all black holes formed in a galaxy become ultra-luminous X-ray sources with luminosity $\ge 10^{39}$ erg/s, explaining the observed population of ULXs. 

Interestingly, only $f_{\rm LMXB}\sim 10^{-6}$ of compact objects become X-ray bright in LMXBs \citep{mineo12}. This is another manifestation of the fact that LMXBs are extremely rare objects and may be explained by the high probability of disruption of the binary system with a low mass companion in the course of the supernova explosion.

These numbers provide valuable input for calibration of the population synthesis models.

\subsection{Connection to LIGO-Virgo sources}

X-ray binaries are one of the few easily observable phases in the evolution of binary stellar systems. Furthermore, since the onset of mass transfer requires small orbital separation, a fraction of X-ray binaries is expected to become binary compact object systems which will merge within a Hubble time producing gravitational-wave sources \citep[e.g.][]{Marchant2017}.  Therefore, the study of X-ray binaries is inextricably linked to the study of the gravitational-wave sources:  X-ray binaries provide information on the demographics of the compact object populations \citep[e.g.][]{Farr2016} and constraints on X-ray binary formation and evolution models. 
Furthermore, joint study of the compact object populations inferred from gravitational-wave obervations and X-ray binaries will provide a more complete picture of their overall mass spectrum and spin distribution \citep[e.g.][]{Fishbach2022}.

\section{Cosmic evolution of X-ray binaries  and their contribution to CXB}

\label{sec:evolution}

\subsection{Contribution of X-ray binaries to cosmic X-ray background}

Knowing how stellar mass was built in the Universe, a natural question to ask is how much X-ray binaries contribute to cosmic X-ray background (CXB). In order to answer this question, \citet{dijkstra12} combined the local   $L_X-{\rm SFR}$ and $L_X-{\rm M_*}$ relations with the cosmic star-formation history. They found that  star-forming galaxies contribute $\sim 5-15$ per cent to soft ($1-2$ keV) CXB and $\sim 1-20$ per cent to the hard band ($2-10$ keV) CXB.  The main source of uncertainty in these estimate was associated with the uncertainty in the spectra of ULXs for which \cite{dijkstra12} allowed a conservatively broad range of photon index values $\Gamma=1-3$. Assuming a more narrow interval of $\Gamma=1.7-2.0$, contribution of star-forming galaxies to the soft CXB can be limited to $\approx 8-13$ per cent.  The contribution to the CXB in the hard band is uncertain mostly because of a more uncertain K-correction at the corresponding high energies. For the parameters of the Chandra Deep Field North, they found that galaxies whose individual observed flux is below the detection threshold in the Chandra Deep Field North (CDF-N), can fully account for the unresolved part of the CXB in the soft band. This conclusion is insensitive to details in the model as long as the photon index, averaged over the entire population of X-ray- emitting star-forming galaxies, is $\Gamma<2$, which corresponds to a very reasonable range given the existing observational constraints on $\Gamma$. 

The tightness of remaining unresolved CXB permits one to constrain the evolution of the $L_X-{\rm SFR}$ relation with redshift. When it is parameterised  as $L_X/{\rm SFR}=A(1+z)^b$,  the unresolved soft CXB requires that $b<1.6$ $(3\sigma)$ \citep{dijkstra12}.

Due to much lower formation efficiency (Section \ref{sec:xrb_freqs}), the contribution to CXB  of LMXBs is at least an order of magnitude smaller \citep{dijkstra12}.

\subsection{X-ray investigations of cosmologically distant galaxies}

With the advent of deep extragalactic \xray\ surveys \citep[see, e.g., review by][]{brandt15}, it is now possible to
place meaningful observational constraints on the cosmic evolution of \xray\
binary populations.  The first deep ($\approx$1~Ms depth) \chandra\ surveys in
the \chandra\ Deep Field-North \citep[CDF-N;][]{hornschemeier00,brandt01}
and CDF-South \citep[CDF-S;][]{giacconi01} revealed substantial numbers
of ``normal'' galaxies at $z \simlt 1.5$ with \xray-to-optical flux ratios
that were consistent with being powered primarily by XRBs and hot gas, without
substantial contributions from active galactic nuclei (AGN).  These discoveries
gave way to the new and active field of \xray\ studies of normal galaxies at
cosmologically significant distances.  Due to the dominance of XRB emission in
normal galaxies at rest-frame wavelengths $>$1--2~keV, combined with the
redshifting of rest-frame soft emission out of the \chandra\ observed frame,
\xray\ emission detected in these distant normal galaxies is expected to primarily
trace XRB populations.  As the Chandra Deep Fields (CDFs) and additional survey fields accumulated
\xray\ depth and multiwavelength data, the insights on the cosmic evolution
of XRB populations in galaxies expanded.

Studies of \xray\ detected normal-galaxy samples in the CDFs have revealed that
the galaxy XLF undergoes positive redshift
evolution from the local universe to $z \approx 1.5$ \citep[e.g.,][]{ptak07,tzanavaris08}.
The evolution is
primarily driven by late-type galaxy populations, which show luminosity
evolution of the XLF at the $(1+z)^{0.4-3.4}$ level.  The XLFs of early-type
galaxies may also evolve with redshift, however, the relatively small numbers
of early-type galaxies detected in the CDFs yield weak constraints on XLF
evolution.  The overall normal galaxy XLF evolution is primarily driven by the
rising cosmic star-formation rate density from $z =$~0--1.5 \citep[see, e.g.,]
[for a review]{madau14}, and scaling-relation investigations show that
the \xray\ detected galaxy population follows the local $L_{\rm X}$/SFR
correlation within the relations scatter \citep[e.g.,][]{mineo14a}.  However,
thus far, X-ray detected normal galaxies in deep surveys number in the hundreds
of objects, which represent only a small fraction (as low as 1\% for {\it
Hubble}-detected sources) of the galaxy population that is known to be present
in these fields.  As such, \xray\ detected normal-galaxy population studies
suffer from significant selection biases.  

\begin{figure*}[t]
\centerline{
\includegraphics[width=\textwidth]{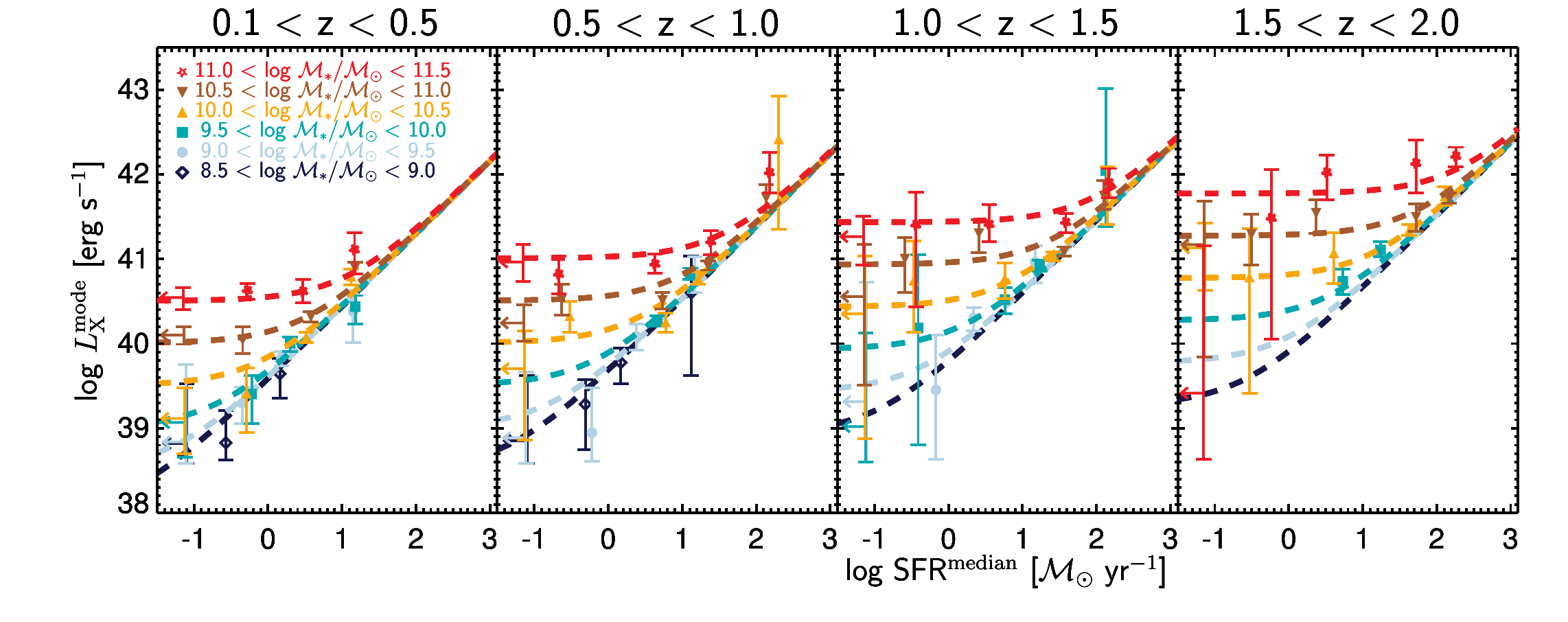}
}
\caption{Most probable X-ray luminosity for galaxy samples selected in bins of redshift, SFR, and $M_\star$, based on analysis of samples in the CANDELS survey fields \citep[Fig~7 of][]{aird17}.  These data show a rise in $L_X$ with redshift, after controlling for SFR and $M_\star$, illustrating the positive evolution of X-ray emission from HMXBs ($L_{\rm X}$(HMXB)/SFR) and LMXBs ($L_{\rm X}$(LMXB)/$M_\star$) with increasing redshift.
}
\label{fig:cdf_scale}
\end{figure*}

To investigate \xray\ emission from more representative populations of galaxies, and
scaling relations between \xray\ binary population luminosity and galaxy
properties, \xray\ stacking of optically-selected galaxies has often been
employed.  Early stacking efforts initially showed that X-ray properties of
normal galaxies were nearly consistent with basic local scaling relations
(e.g., $L_{\rm X}$/SFR and $L_{\rm X}$/$M_\star$) out to $z \approx 3$ once
corrections for optical extinction and $L_{\rm opt}$ to SFR were accounted for
\citep[][]{hornschemeier02,basu13,symeonidis14}.  However, as multiwavelength
data sets expanded and \chandra\ depths and coverage increased, significant
positive redshift evolution was detected in the HMXB and LMXB luminosity scalings with
SFR and $M_\star$, respectively, out to $z \approx 2.5$ (i.e., to a cosmic
lookback time of $\approx$11~Gyr).  The redshift evolution of these scaling
relations has been found to roughly follow $L_{\rm X}$(HMXB)/SFR~$\propto
(1+z)$ and $L_{\rm X}$(LMXB)/$M_\star$~$\propto (1+z)^{2-3}$ \citep[e.g.,][]{lehmer16,aird17,fornasini20}, in good consistency with the CXB based constraints \cite{dijkstra12}.  Figure~\ref{fig:cdf_scale} illustrates constraints from \citet{aird17}, which show that the X-ray luminosity of typical galaxies rises with increasing redshift for fixed stellar mass and SFR.

\subsection{Drivers of the Redshift Evolution of X-ray Binary Populations}

The combination of detailed XRB population synthesis models and cosmological
models of evolving galaxy populations provided a framework to construct models
of the evolution of \xray\ emission from XRBs and their scaling relations with
physical properties.  \citet{fragos13} used {\ttfamily Startrack} XRB
population synthesis models \citep{belczynski08} and simulated galaxy
models from the {\ttfamily Millenium~II} simulation \citep{guo11} to track
the XRB emission throughout the Universe from $z \approx 20$ to the present
day.  These models accounted for evolution in star-formation activity, stellar
masses, stellar ages, and metallicities and identified best models that
simultaneously reproduced local (i.e., $z=0$) $L_{\rm X}$(HMXB)/SFR and $L_{\rm
X}$(LMXB)/$M_\star$ scaling relations.  As a byproduct, these models provided
predictions for the redshift evolution of the scaling relations and the cosmic
\xray\ emissivity from XRB populations.  

The predicted scaling relation evolution from the best Fragos \etal\ models was
found to be similar to that observed empirically in the CDFs, and the models
provided physical insight for the drivers of this evolution.  The rise of
$L_{\rm X}$(HMXB)/SFR with redshift was expected to be driven by a
corresponding decline in metallicity.  Stellar wind mass loss is expected to
increase with metal content, and in the context of XRBs, relatively low
metallicity systems are expected to lose less mass and angular momentum from
stellar winds, allowing the binary orbits to remain relatively tight and yield
larger mass-transfer rates and higher \xray\ luminosities than relatively high
metallicity systems.  For LMXBs, the increase of $L_{\rm X}$(LMXB)/$M_\star$
with increasing redshift was predicted theoretically as a result of the LMXB
donor stars shifting to higher mass objects and higher mass-transfer rates as
the ages of the stellar populations decline with increasing redshift. 

\begin{figure*}[t]
\centerline{
\includegraphics[width=0.5\textwidth]{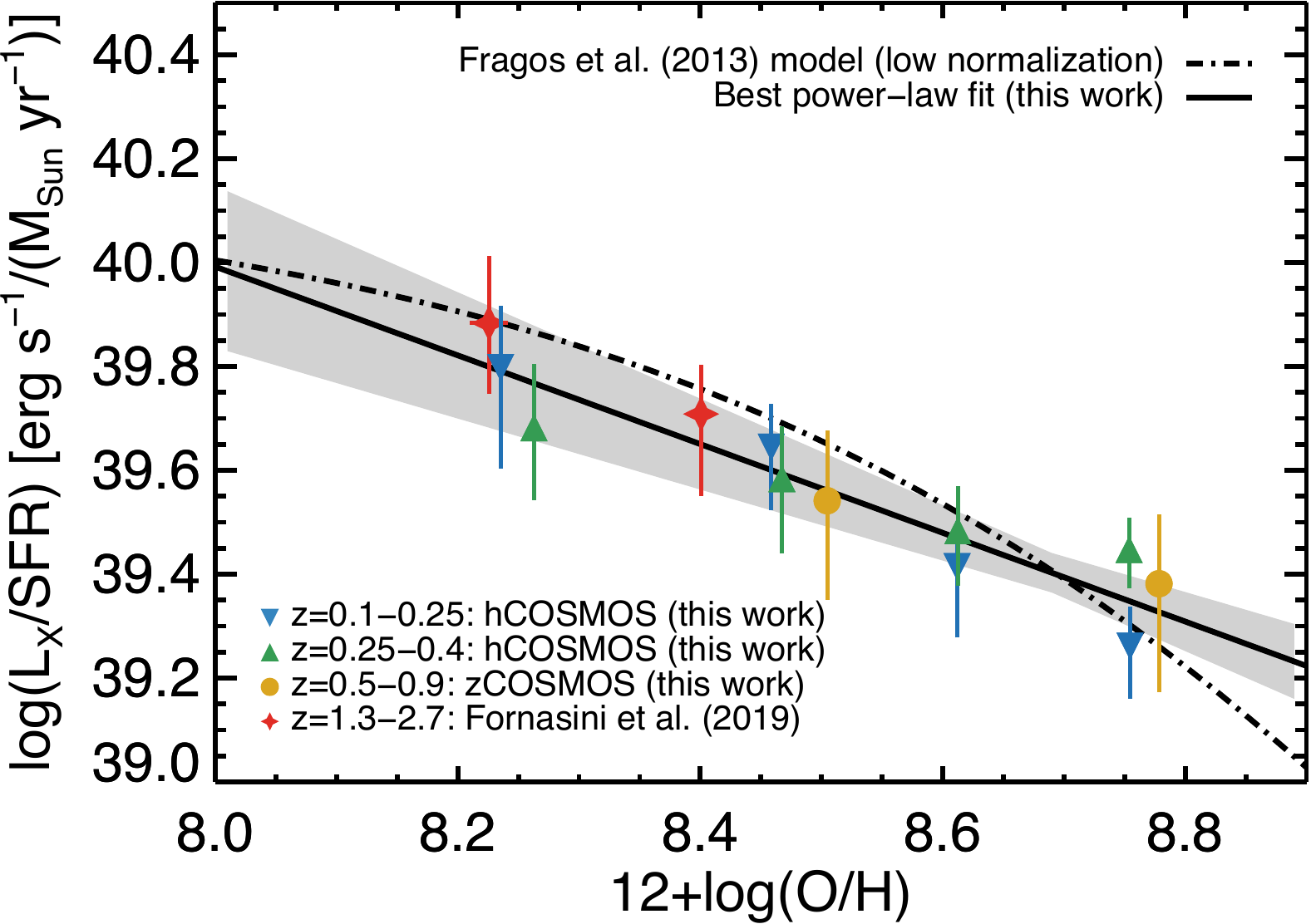}
\includegraphics[width=0.5\textwidth]{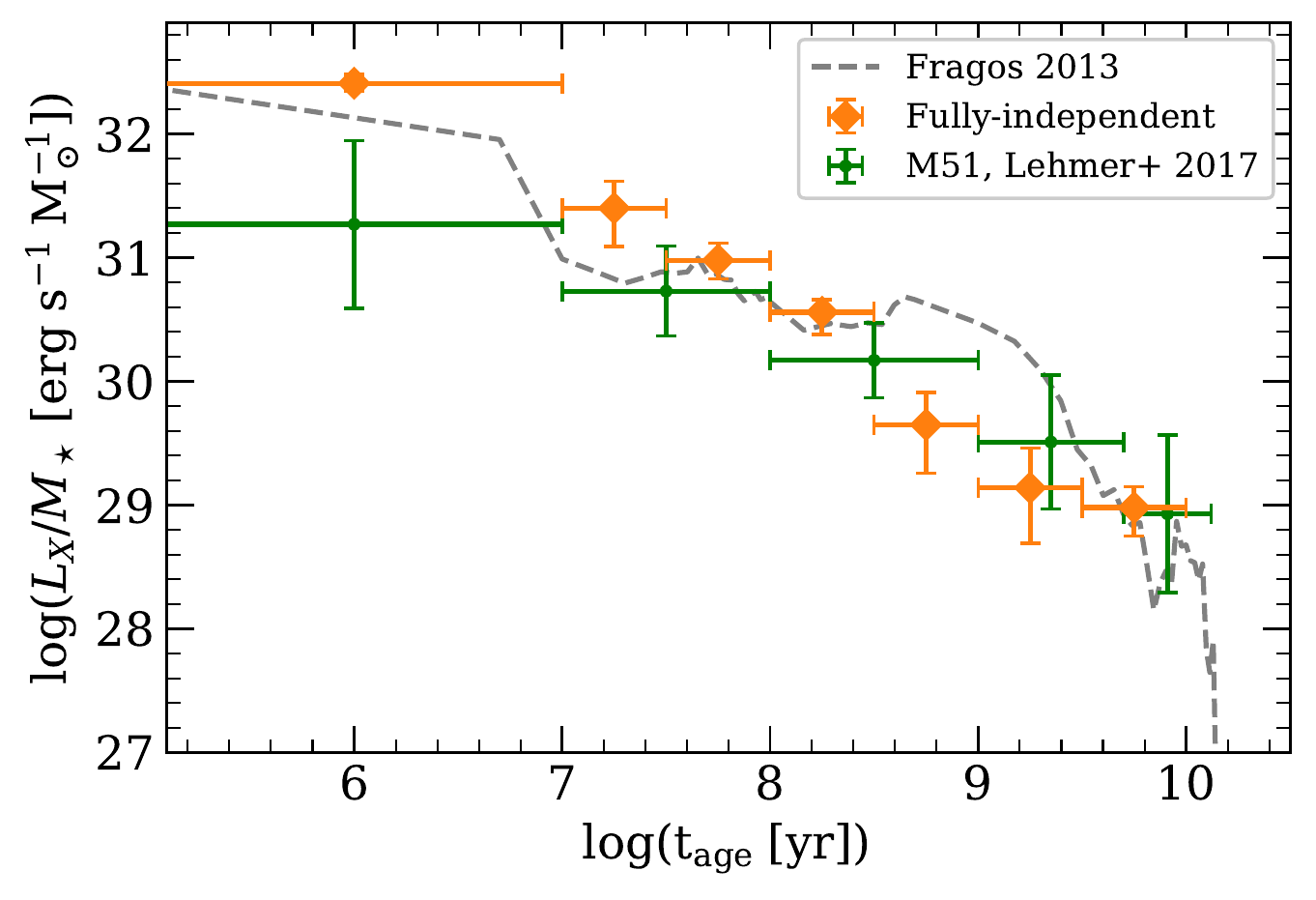}
}
\caption{
({\it Left}) The X-ray luminosity per SFR ($L_{\rm X}$/SFR) versus gas-phase metallicity (12 + $\log({\rm O/H})$) for stacked samples of high specific star-formation rate galaxies at $z \approx$~0.1--2.6 in the CDF and COSMOS survey fields.  The X-ray emission for these galaxies is expected to be dominated by young populations of HMXBs, and a clear anticorrelation is observed (black curve indicates best-fit relation), consistent with theoretical models (the dot-dashed curve from \citet{fragos13}).  Adapted from \cite{fornasini20}.
({\it Right}) Average age-dependence of the 2--10~keV luminosity per stellar mass ($L_{\rm X}/M_\star$) for 344 galaxies in the CDFs.  A decline of $\approx$3 orders of magnitude is observed from 10~Myr to 10~Gyr as XRB populations become less luminous and powered by accretion from increasingly lower mass stars. Figure modified from its original version in \citep{gilbertson21}.
}
\label{fig:hz_scaling}
\end{figure*}

\subsection{Recent Constraints on X-ray Evolution of Galaxies}

\label{sec:evol}

While theoretical models suggest that both metallicity and stellar age
evolution are responsible for the observed evolution of XRB scaling relations,
more direct empirical evidence has only recently supported these suggestions.
For example, \citep{fornasini19, fornasini20} investigated HMXB-dominant galaxies
(high SFR/$M_\star$) at $z \approx$~0.1--2.6, located in the COSMOS and CDF-S
fields, that had gas-phase metallicity measurements via strong emission-line
indicators.  They divided their galaxy samples into bins of redshift and
gas-phase metallicity and used \xray\ stacking to show that the mean $L_{\rm
X}$(HMXB)/SFR ratio declined with increasing metallicity in a single relation
that is consistent with local-galaxy $L_{\rm X}$(HMXB)-SFR-$Z$ relations \citep[e.g.,][]{brorby16,lehmer21} -- see left-panel of Figure~\ref{fig:hz_scaling}.  To investigate age evolution of XRB
populations, \citep{gilbertson21} measured star-formation histories of
galaxies in the CDFs, and used a statistical method to construct a model of
$L_{\rm X}/M_\star$ versus age consistent with the galaxy X-ray counts.  A decline of three orders of magnitude was observed for
$L_{\rm X}/M_\star$ from $\approx$10~Myr to $\approx$10~Gyr, mainly consistent
with expectations from the \citet{fragos13} models (right-panel of Figure~\ref{fig:hz_scaling}).

\subsection{Contribution to (pre)heating of IGM}

\label{sec:preheating}

The large mean-free path of  X-ray photons suggests that they can penetrate a larger volume of the interstellar medium around the X-ray source. This becomes particularly important in the early Universe ($z>10$), where they may influence a larger volume around the primordial galaxies than the ultraviolet photons associated with the first Population-III stars \citep[e.g.][and references therein]{madau17}. This results in heating of the primordial intergalactic medium even before the epoch of reionization, which has important implications for the subsequent galaxy formation  \citep[e.g.][]{Artale2015}.

Extrapolation of the best XRB population synthesis models to $z \approx$~3--20,
where only weak empirical constraints are currently available, indicate that
X-ray emissivity of the Universe from HMXBs is likely to exceed AGN at $z
\simgt$~6--8 \citep[e.g.,][]{fragos13,madau17}.  At these
redshifts, $L_{\rm X}$(HMXB)/SFR is expected to be elevated over the local
relation by a factor of $\sim$10 due to differences in metallicity.  As an
added result, the escaping radiation at low energies $\simlt$2~keV may be
further enhanced due to the lack of metal absorption edges that significantly
impact the optical depth at these energies \citep[e.g.,][]{das17}.  As a
consequence of these effects, HMXB populations are of particular interest as
having a potentially significant role in heating of the intergalactic medium at
$z \simgt 10$.  The impact of this heating is expected to be imprinted at these
redshifts on the cosmic 21-cm brightness temperature relative to the cosmic
microwave background, and numerous efforts are underway to directly observe
these signatures.  For example, the Hydrogen Epoch of Reionization Array \citep[HERA; e.g.,][]{deboer17} and Square Kilometre Array \citep[SKA; e.g.,][]{mellema13}, are predicted to directly constrain the 21-cm signal
over a wider range of redshift and will enable constraints on $L_{\rm
X}$(HMXB)/SFR associated with the galaxy populations there \citep[see, e.g.,][for first results]{hera21}.

\section{Concluding remarks}

Sub-arcsecond angular resolution of Chandra observatory led to a quantum leap in our understanding of  populations of X-ray binaries in external galaxies, their content, evolution and scaling with fundamental parameters of galaxies. In over 20 years of operation in space, Chandra observed hundreds of galaxies of various morphological types, ages and metallicities. X-ray luminosity functions of compact sources in young and old galaxies to the meaningful depth have been obtained for many dozens of galaxies, providing the ultimate proof of their nature as  high- and low-mass X-ray binaries. Their spatial distributions have been obtained and compared with the distributions of various tracers, confirming that different formation channels are in place, primordial and dynamical. These observations provided wealth of information for calibration and verification of population synthesis models which in turn gave valuable feedback to the theories of stellar and binary evolution.  Deep Chandra fields permitted to study collective properties of high-mass X-ray binaries in distant galaxies, located at cosmological redshifts, to study their evolution and to make further comparisons with  binary population modelling. These studies revealed an unanticipated role of star-forming galaxies and their X-ray binaries in preheating the inter-galactic medium in early Universe and in shaping the Cosmic X-ray background, and also led to the proposition of a new independent method to measure star-formation rate in (distant) galaxies.

Along with these remarkable advances, many unanswered questions still remain. The list of outstanding goals for future studies includes, to mention a few: the redshift evolution of X-ray binaries, the nature and formation channels of ultra-luminous X-ray sources and their  connection to LIGO-Virgo sources, the maximum mass of "stellar mass" black holes and role of the intermediate mass ones,  detailed understanding of dynamical formation of LMXBs in globular clusters and galactic nuclei and their metallicity dependence  and seeding of field populations, the origin of the HMXB XLF, which  maintains same slope over five orders of magnitude in luminosity, with only moderate deviations from the power law. On the other end of the luminosity range are fainter sources such as cataclismic variables which extragalactic populations are yet to be explored. There are still many complexities in the data which we do not quite understand, but  existence of many others is yet to be recognised.

The  progress of observational capabilities of modern astronomy, anticipated in the coming years and planned for the more distant future, will help to answer these questions and will inevitably raise the new ones.
Eagerly awaited are the results from the SRG all-sky survey which is more than a half a way through. Its eROSITA telescope, although lacking the angular resolution of Chandra, will survey the entire sky eight times. It will detect of the order of $\sim 10^4$ normal galaxies of all morphological types, fully sampling the parameter space of normal galaxies. Critical role in many studies is played by observations at other wavelength, in particular, optical and infra-red. These will be advanced with the start of operations of the Vera C. Rubin Observatory and Euclid satellite. Smaller samples of carefully selected galaxies will be observed by the James Webb Space telescope.  At the X-ray wavelengths, the major new thrust will be given by Athena and Lynx X-ray observatories. Notably, the Lynx Observatory will deliver the angular resolution of Chandra but at $\sim 50$ times higher throughput.

\end{document}